\documentclass[11pt]{article}
\usepackage[DIV13]{typearea}
\usepackage{amsmath}
\usepackage{amsfonts}
\usepackage{amssymb}
\usepackage[titletoc]{appendix}
\usepackage{array}
\usepackage{bbold} 
\usepackage{bbm}
\usepackage{booktabs}
 \usepackage{cancel}
 \usepackage{cite} 
\usepackage[usenames,dvipsnames]{color}
\usepackage{epsfig}
\usepackage{fancyhdr}
\usepackage[T1]{fontenc}
\usepackage{framed}
\usepackage[left=.9in, right=.9in]{geometry}
\usepackage{graphicx}
\usepackage{hhline} 
\RequirePackage[colorlinks=true,urlcolor=blue,anchorcolor=blue,citecolor=blue,filecolor=blue,
               linkcolor=blue,menucolor=blue,linktocpage=true,pdfproducer=medialab]{hyperref}
\usepackage[utf8]{inputenc}
\usepackage{mathrsfs}
\usepackage{mathtools}
\usepackage{dsfont}
\usepackage{multicol}
\usepackage{multirow}
\usepackage{indentfirst}
\usepackage{longtable}
 \usepackage{lscape}
\usepackage{pbox}
\usepackage{pdfpages}
\usepackage{rotating}
\usepackage{slashed}
\usepackage[DIV13]{typearea}
\usepackage[normalem]{ulem}
\usepackage{units}
\usepackage{enumitem}
\usepackage{soul}
\usepackage{placeins}
\usepackage{tikz}

\usepackage{catchfile}
\newcommand{\getenv}[2][]{%
  \CatchFileEdef{\temp}{"|kpsewhich --var-value #2"}{}%
  \if\relax\detokenize{#1}\relax\temp\else\let#1\temp\fi
}

\allowdisplaybreaks
\renewcommand{\to}{\rightarrow}

\newcommand{\hc}{\mathrm{h.c.}}

\newcommand{\beq}{\begin{equation}}
\newcommand{\eeq}{\end{equation}}
\newcommand{\bea}{\begin{eqnarray}}
\newcommand{\eea}{\end{eqnarray}}
\renewcommand{\[}{\begin{equation}}
\renewcommand{\]}{\end{equation}}
\definecolor{orange}{rgb}{1,0.5,0}

\newcommand{\blue}[1]{\color{blue} #1 \color{black}}

\newcounter{diagram}

\newcommand{\email}[1]{\href{mailto:#1}{\tt #1}}

\renewcommand{\to}{\rightarrow}

\renewcommand{\to}{\longrightarrow}

\renewcommand{\[}{\begin{equation}}
\renewcommand{\]}{\end{equation}}
\newcommand{\bmat}{\begin{pmatrix}}
\newcommand{\emat}{\end{pmatrix}}

\newcommand{\gamup}{\gamma^\mu}

\newcommand{\dmup}{D^\mu}

\definecolor{green}{rgb}{0.13, 0.55, 0.13}

\newcolumntype{C}[1]{>{\centering\let\newline\\\arraybackslash\hspace{0pt}}m{#1}}

\begin{document}
\vspace*{-1cm}
\phantom{hep-ph/***} 
{\flushleft
{\blue{IFT-UAM/CSIC-18-050}}
\hfill{\blue{FTUAM-18-13}}
}
\vskip 1.5cm
\begin{center}
{\LARGE Color Unified Dynamical Axion\\[0.5cm] }
\vskip .3cm
\end{center}
\vskip 0.5  cm
\begin{center}
{\large M.K.~Gaillard}~$^{a)}$,
{\large M.B.~Gavela}~$^{b)}$,
{\large R.~Houtz}~$^{b)}$, 
{\large P.~Quilez}~$^{b)}$
{\large R.~del Rey}~$^{b)}$,
\\
\vskip .7cm
{\footnotesize
$^{a)}$~
Department of Physics and Theoretical Physics Group, Lawrence Berkeley Laboratory,
University of California, Berkeley, California 94720\\
\vskip .1cm
$^{b)}$~
Departamento de F\'isica Te\'orica and Instituto de F\'{\i}sica Te\'orica, IFT-UAM/CSIC,\\
Universidad Aut\'onoma de Madrid, Cantoblanco, 28049, Madrid, Spain
\vskip .5cm
\begin{minipage}[l]{.9\textwidth}
\begin{center} 
\textit{E-mail:} 
\email{mkgaillard@lbl.gov},
\email{belen.gavela@uam.es},
\email{rachel.houtz@uam.es},
\email{rocio.rey@uam.es},
\email{pablo.quilez@uam.es}
\end{center}
\end{minipage}
}
\end{center}

\begin{abstract}
We consider an enlarged color sector which solves the strong CP
problem via new massless fermions. The spontaneous breaking of a
unified color group into QCD and another confining group provides a
source of naturally large axion mass $m_a$ due to small size
instantons. This extra source of axion mass respects automatically the
alignment of the vacuum, ensuring a low-energy CP-conserving vacuum.
The mechanism does not appeal to a $Z_2$ ``mirror'' copy of the SM, nor
does it require any fine-tuning of the axion-related couplings at the
unification scale.  There is no light axion and uncharacteristically
the lighter spectrum contains instead sterile fermions.  The axion
scale $f_a$ can be naturally brought down to a few TeV, with an exotic
spectrum of colored pseudoscalars lighter than this scale, 
observable at colliders exclusively via strong interactions. The
$\{m_a, f_a\}$ parameter space which allows a solution of  the strong CP
problem is thus enlarged well beyond that of invisible axion models.

\end{abstract}

\vskip 1cm

\pagebreak
\tableofcontents

\pagebreak

%
%
\section{Introduction}
\label{Sect:Intro}

Phenomenological analyses based on chiral perturbation theory and supported by lattice computations indicate that all Standard Model (SM) quarks have non-zero masses. This disfavors the solution to the strong CP problem via one massless SM quark, which automatically guarantees a $U(1)$ axial invariance at the classical level.  The interesting possibility of having a massless up quark in the microscopic theory which appears as massive at QCD scales  due to non-perturbative instanton contributions~\cite{Georgi:1981be} does not seem to be realized in nature, even if this option is not completely excluded~\cite{Dine:2014dga, Choi:1988sy, Kaplan:1986ru, Frison:2016rnq, Aoki:2013ldr, Aoki:2016frl}.

It is still possible to solve the strong CP problem using massless fermions if the SM up quark is massive.
The idea is to enlarge the SM gauge group with a new confining sector~\cite{Weinberg:1975gm,Susskind:1978ms,Dimopoulos:1981xc}, whose scale is much larger than that of the QCD group, $SU(3)_c$.  Extra massless quarks charged under both QCD and the new confining sector may realistically solve the 
problem~\cite{Choi:1985cb}. A new spectrum of confined states results.

 The term ``axion'' denotes any  (pseudo)Goldstone boson (pGB) of a global chiral $U(1)$ symmetry which is exact at the classical level but has anomalous couplings to the field strength of a confining group.~\footnote{The SM $\eta'$ is excluded from this definition since the $U(1)_A$ symmetry associated to it is broken by the non-zero quark masses.}  Axions are characteristic of solutions to the strong CP problem based on an anomalous $U(1)$ axial symmetry, usually called Peccei-Quinn (PQ) symmetry~\cite{Peccei:1977hh}. When the number of axions in a given theory outnumbers the total number of distinct instanton-induced scales other than QCD to which they couple,  one (or more) light axions remain.  Axions which are not elementary but composed of fermions are typical of theories with an enlarged confining sector and are often referred to as ``dynamical'' or composite axions.
  Very heavy dynamical axions made out  of massless fermions will typically acquire the bulk of their mass from the largest instanton-induced scale $\Lambda$ to which they exhibit anomalous couplings,
  \begin{equation}
 m_a^2 f_a^2 \sim \Lambda^4\,. 
 \label{heavyaxion}
 \end{equation}
  For a very light axion coupled only to QCD, the mixing with  the $\eta'$ pseudoscalar is  relevant instead, and that axion obeys the usual relation~\cite{Weinberg:1977ma,Georgi:1986df}
 \begin{equation}
 m_a^2 f_a^2 \sim m_\pi^2 f_\pi^2\,\frac{m_u\,m_d}{(m_u+m_d)^2}\,,
 \label{invisiblesaxion}
 \end{equation}
where 
 $m_\pi, f_\pi, m_u, m_d$ denote 
 the pion mass and coupling constant, and the up  and down quark masses, respectively.

The first step in the direction  of solving the strong CP problem with exotic massless quarks  
was the proposal by 
 K. Choi and E. Kim \cite{Choi:1985cb}  to enlarge the confining gauge sector of the SM to $SU(3)_c \times SU(\tilde{N})$, with the latter having a scale larger than that of 
 QCD, $\tilde{\Lambda} \gg \Lambda_{\rm{QCD}}$.  Two confined charges would then exist in nature, color and  {\it axicolor } respectively, and correspondingly two distinct sources of instanton potentials.  
 A massless  color-triplet quark $Q$,  charged also under axicolor and singlet under the SM electroweak symmetry,  would solve the QCD strong CP problem. The fact that the axicolor scale is  very large  would explain the non-observation of exotic bound states at low energies. An issue arises because  there are now two potentially harmful vacuum angles  to absorb:  $\theta_c$ of QCD and $\tilde{\theta}$. Only one combination of them would be redefined away by a chiral rotation of $Q$. 
 This was easily remedied by adding a second exotic quark $\chi$ charged only under $SU(\tilde{N})$. In the limit of vanishing QCD coupling, the $SU(\tilde{N})$ sector  described four flavors which seed two singlet pseudoscalars with anomalous couplings: two dynamical axions. Finally,  taking into account the SM quark sector and thus the SM $\eta'$,  three flavor-singlet pseudoscalars result in the low-energy spectrum 
 for only two instanton sources of masses:  the $\eta'$,  a very heavy axion with mass $\sim \tilde{\Lambda}$ and a second axion almost massless and obeying 
  Eq.~(\ref{invisiblesaxion}).
Because of this last axion,
  the axicolor construction can be seen as an ultraviolet dynamical completion of the invisible axion paradigm. As usual, a very large $f_a$ scale is required to be orders of magnitude larger than the electroweak (EW) one, albeit with the advantage of being free from scalar potential fine-tunings.

 In a different and recent attempt~\cite{Dimopoulos:2016lvn} to solve the strong CP problem with extra massless quarks,  the same $SU(3)_c\times SU(\tilde{N})$ confining sector is considered. No light axion remains in the low-energy spectrum, though, as only two pseudoscalar mesons are present which couple to the two anomalous currents: the customary $\eta'$ meson and one axion. This is achieved by assuming only one exotic massless quark $Q$ instead of two. In the limit of vanishing QCD coupling, the $SU(\tilde{N})$ sector then describes only three flavors, resulting in only one gauge singlet pseudoscalar with anomalous $SU(\tilde{N})$ couplings: a dynamical axion. 
 Both the $\eta'$
and the axion thus acquire a mass,  and the axion mass is induced
by $SU(\tilde{N})$ instantons, $m_a, f_a \sim \tilde{\Lambda}$.  For such a heavy axion 
 the 
$f_a$ scale can be as low as the TeV range without
incurring unacceptable phenomenological consequences.  The issue of
the two $\theta$ parameters is solved in this proposal by imposing a
discrete $Z_2$ symmetry relating the two sectors. Unfortunately, the
practical implementation of this idea requires a complete
$Z_2$-``mirror'' copy of the SM. The $Z_2$ symmetry is explicitly
broken by a scalar potential which gives the second Higgs field a very
large vacuum expectation value, e.g. $10^{14}$ GeV, in order to sufficiently modify 
the running of the two confining scales. This is overall a quite
complex, tuned, and large structure.

 Even more recently, the SM massless quark avenue has been revisited in a theory that involves a product of 
 $SU(3)$ groups which break spontaneously to QCD~\cite{Agrawal:2017evu}. A very interesting aspect developed by the same authors in a previous work~\cite{Agrawal:2017ksf} is the impact of small size instantons  of the group which undergoes spontaneous symmetry breaking. These instantons are shown to provide a possible extra source of large masses for the putative axions of the model.

We will develop in what follows a new solution to the strong CP problem via massless fermions, in which the issue of the  different $\theta$ parameters that arise in the presence of two or more confining groups  is solved via color unification.  Color unification with massless quarks is attempted here for the first time. This path is an alternative to the axicolor-type constructions and will lead to different phenomenology.   
QCD will be unified with another confining sector singlet under the electroweak gauge symmetry. The color unified theory (CUT) breaks spontaneously to QCD and another confining group.
 The small-size instantons of the unified color group provide an extra source of high masses for the axions of the theory, and it will be shown that no axion remains at low scales. The exotic low-energy spectrum is instead fermionic. 
 Furthermore, it will be shown that interesting new phenomenological signals can be explored at colliders.  The complete ultraviolet completion of this idea will be developed, implementing two different scenarios: in one of them the two resulting heavy axions are dynamical, while in the other  one axion  is elementary.

The structure of the paper can be easily inferred from the Table of Contents.

\newpage

\section{ SU(6) Color Unification }

We propose a scenario in which QCD is unified with another confining group into  $SU(6)$, and a single, strictly massless $SU(6)$ fermion rotates away simultaneously all $\theta$ parameters.   The unification path in the context of an extended strong sector to solve the strong CP problem was first proposed by Rubakov long ago,~\cite{Rubakov:1997vp} in a Grand Unification construction that relied on traditional models {\it \`a la} DFSZ~\cite{Zhitnitsky:1980tq,Dine:1981rt} with massive exotic fields, and  required a $Z_2$ mirror copy of the complete SM field content. Another recent  attempt~\cite{Gherghetta:2016fhp} using unification ideas also relied on massive exotic fermions {\it \`a la} DFSZ.    
 Here we instead consider color unification in the presence of massless fermions. 
 The massless $SU(6)$ fermion belongs to the 20 representation of the $SU(6)$ CUT, having a definite chirality (e.g. left-handed) while  being a singlet of the SM $SU(2)_L\times U(1)_Y$ gauge symmetry:
\begin{table}[h!]
\begin{align*}
\begin{array}{c|c|c|c}
	& SU(6)	& SU(2)_L	
								& U(1)	\\
\hline
\Psi_L	& 20		& 1					& 0		
\end{array}
\end{align*}
\label{massless-above}
\caption{The massless fermion sector of the $SU(6)$ construction above the unification scale.}
\end{table}

At a color unification scale $\Lambda_{\rm{CUT}}$ much higher than the EW one,  the $SU(6)$ group breaks into
\begin{equation}
SU(6)\xrightarrow{\Lambda_{\rm{CUT}}} SU(3)_c\,\times SU(\tilde{3})\times U(1)\,.
\label{SU6onlybreak}
\end{equation}
The parameters $\theta_c$  of $SU(3)_c$ and  $\tilde{\theta}$ of $SU(\tilde{3})$ are necessarily equal and unphysical down to the unification scale, and will  remain so even below the unification scale as long as $\Psi$ remains massless, protected by chiral symmetry. 
 Under  spontaneous symmetry breaking of the CUT symmetry, 
 $\Psi$  decomposes as
\begin{equation}
\Psi_L(20)= (1, 1)(-3)_L\,\,+\, (1, 1)\,(+3)_L +\, (3, \bar3)\,(-1)_L+\, (\bar3, 3)({+1})_L\,,
\label{Psidecomposition}
\end{equation}
where the charges under the $U(1)$ group in Eq.~(\ref{SU6onlybreak}) are shown in parenthesis for completeness.  If the components of the $\Psi_L$ field are to remain massless under the CUT scale, $SU(\tilde{3})$ must confine. The two confining scales $\Lambda_{\text{QCD}}$ and $\tilde{\Lambda}$  need to be separated with  $\tilde{\Lambda}\gg \Lambda_{\text{QCD}}$, as no bound states are observed other than those compatible with QCD. 
The 20-dimensional representation is thus advantageous because all its components charged under QCD are also charged under $SU(\tilde{3})$, and so will form bound states at the higher scale $\tilde{\Lambda}$. This representation is also pseudo-real, and so the theory is anomaly free. The non-trivial issue of how to separate $\tilde{\Lambda}$ and $\Lambda_{\text{QCD}}$ is discussed further below.

\begin{table}[h!]
\begin{align*}
\begin{array}{c|c|c}
	& SU(3)_c	& SU(\tilde{3})	\\
\hline
\psi_{L}	& \Box		& \bar\Box	\\
\psi_{L}^c	& \bar\Box		& \Box	\\
2\times\psi_\nu	& 1		& 1	
\end{array}
\end{align*}
\caption{The massless fermion sector of the $SU(6)$ construction below the unification scale. The notation is such that $\psi_L^c\equiv (\psi^c)_L=(\psi_R)^c$. }
\label{massless-fermions}
\end{table}
The  colored-axicolored massless fermions in  Eq.~(\ref{Psidecomposition}) will be denoted $\psi_{L,R}$, see Tab.~\ref{massless-fermions}, while $\psi_\nu$ will refer to the singlet massless fermions 
    to convey that they act like sterile\footnote{By ``sterile fermion'' is meant any fermion which is not charged under the SM gauge group.} neutrinos. The $\psi_\nu$ fields only connect  to the other fields through the unified  strong forces, and thus their couplings to the visible universe will be safely suppressed by $\Lambda_\text{CUT}$,
provided the $U(1)$ gauge group in (\ref{SU6onlybreak}) is also
broken near that scale.\footnote{This breaking will become manifest in the next section.}  
    
    $SU(6)$ color unification is thus a successful path to solve the strong CP problem, and this fact will remain at the heart of the  developments in this paper. The remaining problem is to obtain a low-energy spectrum which is fully compatible with observations.

  \noindent \subsubsection*{ The SM fermions}
   \vspace{-0.3cm}
 Because of color unification, the SM quarks must belong to $SU(6)$ multiplets. The simplest option is to include them in  six-dimensional fundamental representations. For each fermion generation,
  \begin{equation}
  Q_L (6)\,\equiv \, ( q\,,\, \tilde{q})_L\,,\qquad U(6)\,\equiv \, (u\,,\, \tilde{u})_R\,,\qquad D(6)\,\equiv \, (d\,,\, \tilde{d})_R\,\,\,,
 \label{qqtildebis}
  \end{equation}
  where $q_L$, $u_R$ and $d_R$  denote the SM quarks, while their $SU(6)$ partners are signaled by tildes.  
   The $\tilde{q}_L$ fields are necessarily electroweak doublets, and this character turns out to be the major practical issue of this model:
 \begin{itemize}
  \item  Leaving the tilde-quark sector massless but confined is unacceptable,  as the condensate\,---\,assuming chiral symmetry breaking of $SU(\tilde{3})$\,---\,would typically break the SM EW symmetry at the  large $\tilde{\Lambda}$ scale.
  \item Alternatively, giving much larger masses ( $\ge{\tilde \Lambda}$ ) to the tilde quarks is not viable either in this $SU(6)$ setup without spoiling SM quark masses, since they belong to the same multiplet. If a scalar field gave high masses to the tilde quarks~\footnote{Through tuned Yukawa couplings of the tilde-quark sector to an extended scalar sector.} by obtaining a high vacuum expectation value (vev), that scalar field would have to be an $SU(2)_L$ doublet. Then its large vev would spontaneously break SM EW symmetry, giving gigantic masses to the $W$ and $Z$ boson.

  \end{itemize}
  The  main problem of this model is then  the unacceptably light tilde-fermion sector.  We will develop  next an extension whose only purpose is precisely to achieve high masses for the tilde-sector quarks, decoupling them from the low-energy spectrum. By the same token, the necessary separation of $\Lambda_{\rm{QCD}}$ and  a larger confining scale  will naturally follow.\footnote{If $SU(6)$ sufficed to obtain a realistic spectrum, the $SU(\tilde 3)$ group and $\tilde \Lambda$ scale of this section would correspond to those of the axicolor group~\cite{Choi:1985cb} as described in the introduction. The extension of the CUT group will break this direct correspondence, although two confining groups will still be at play. } We will develop in detail two realistic ultraviolet (UV) completions.

\section{ The realistic Color Unified Theory: $SU(6)\times SU(3')$ }
\label{realistic}

 It is necessary to give large masses to the tilde-quark sector without giving masses to the SM quarks, a challenging enterprise as explained above due to the $SU(6)$ unification.  An external mechanism is ideal for this task. 
  The color unified $SU(6)$ group which contains QCD is enlarged via an external non-abelian $SU(3')$ group with  additional fermions charged only under the latter.  In fact,  
all fermions in the theory will be charged  under  only one of the two groups, $SU(6)$ or $SU(3')$. $\Psi_L$ will be thus taken to be a singlet of $SU(3')$ and the same applies to the multiplets in Eq.~(\ref{qqtildebis}) which contain the SM quarks.  
  The two sectors are connected exclusively via a new scalar $\Delta$.  QCD remains a subgroup of $SU(6)$, whose $\theta$-parameter is rotated away by the massless $\Psi_L$ fermion in Eq.~(\ref{Psidecomposition}).     This type of auxiliary extension was suggested in Ref.~\cite{Gherghetta:2016fhp} to give high masses to exotic fermions in a different context. The  field content of our model is summarized in Tab.~\ref{model-matter-compact}, in which all fermions except $\Psi_L$ will become massive.  It is easy to see that the theory with this matter content is anomaly free. The scalar $\Delta$ appearing in the table belongs to the bifundamental  of $SU(6)\times SU(3')$, and its vev  breaks color unification  at a scale $\Lambda_{\rm{CUT}}$, taken to be much larger than all SM scales,  
\vspace{-0.3cm}
\begin{table}
\centering
\begin{tabular}{c|c|c >{\quad}c<{\quad}  >{$}c<{$}|>{$}c<{$}|>{$}c<{$}||>{$}c<{$}|>{$}c<{$}}
					& $SU(6)$				& $SU(3')$				&&			& SU(3)_c		& SU(3)_{\rm{diag}} &SU(2)_L		& U(1)_Y	\\
\cline{1-3}\cline{5-9}
\multirow{2}{*}{$Q_L$}	& \multirow{2}{*}{$ \Box$}	& \multirow{2}{*}{1}		&& {q_L}		& \Box		& 1				& \Box		& \frac16 	\\[-.5ex]
					& 					&					&& \mathbf{\tilde{q}_L}	& 1			& \Box			& \Box		& \frac16	\\
\cline{1-3}\cline{5-9}
 \multirow{2}{*}{$U_L^c$}	& \multirow{2}{*}{$ \bar\Box$}& \multirow{2}{*}{$1$}	&&u_L^c		& \bar{\Box}	& 1				& 1			& -\frac23	\\[-.5ex]
 					&					&					&&\mathbf{\tilde{u}_L^c}	& 1			& \bar{\Box}		& 1			& -\frac23	\\
\cline{1-3}\cline{5-9}		
 \multirow{2}{*}{$D_L^c$}	&  \multirow{2}{*}{$\bar\Box$}&  \multirow{2}{*}{$1$}	&&d_L^c		& \bar{\Box}	& 1				& 1			& \frac13	\\[-.5ex]	
  					&					&					&&\mathbf{\tilde{d}_L^c}	& 1			& \bar{\Box}		& 1			& \frac13	\\
\cline{1-3}\cline{5-9}
 \multirow{3}{*}{$\Psi_L$}	&  \multirow{3}{*}{$ 20$}	&  \multirow{3}{*}{$1$}	&& \psi_{L}	& \Box		& \bar{\Box}		& 1			& 0		\\[-.5ex]
					&					&					&&\psi^c_{L}	& \bar\Box		& {\Box}			& 1			& 0		\\[-.5ex]
					&					&					&&2\times\psi_\nu & 1		& 1 				& 1			& 1		\\
\cline{1-3}\cline{5-9}
$q_L'$				& $1$				& $\bar\Box$			&& \mathbf{q'_L}		& 1			& \bar\Box			& \Box		& -\frac16	\\
\cline{1-3}\cline{5-9}
${u_L'^c}$				& $1$				&$ \Box$			&& \mathbf{u'^c_L}		& 1			& \Box		& 1			& \frac23	\\
\cline{1-3}\cline{5-9}
${d_L'^c}$				& $1$				&$ \Box$					&& \mathbf{d'^c_L}		& 1			& \Box		& 1			& -\frac13	\\
\cline{1-3}\cline{5-9}
$\Delta$				&$ \Box$				&$ \bar \Box	$			&& -			& -			& -				&	1		& 0		\\[-.5ex]
\end{tabular}

\caption{The matter content. The table on the left describes matter above the CUT scale, while the one on the right gives the transformation properties under the gauge groups remaining  after CUT spontaneous breaking. The fermions in bold have masses comparable to $\Lambda_{\rm{CUT}}$ and are integrated out around the CUT scale. The quantum numbers under the EW gauge group correspond both to the high-energy and low-energy fields.}
\label{model-matter-compact}
\end{table}

\begin{equation}
SU(6)\times SU(3')\xrightarrow{\Lambda_{\rm{CUT}}}\,SU(3)_c\,\times SU({3})_{\rm{diag}}\,.
\label{SU6break}
\end{equation}
The fermion quantum numbers under the two resulting groups are also shown in Tab.~\ref{model-matter-compact}. 

A simple CUT-invariant  Yukawa Lagrangian which connects the $SU(6)$ and the auxiliary $SU(3')$ extension reads 
		\begin{align}
	\mathcal{L}\ni \kappa_q \,q_L' \Delta^* Q_L
		+ \kappa_u \, u_L'^c \Delta\, U_L^c 
		+  \kappa_d \, d_L'^c \Delta D_L^c \, +\, \text{h.c.}\,.
		\label{LDelta}
	\end{align}
The CUT symmetry is spontaneously broken upon $\Delta$ taking a vev 
 of the order of the CUT breaking scale $ \Lambda_{\rm{CUT}}$
\begin{align}
	\langle \Delta \rangle =  \Lambda_{\rm{CUT}} \begin{pmatrix} 	0 & 0 & 0 & 1 & 0 & 0 \\
									0 & 0 & 0 & 0 & 1 & 0 \\
	\label{Deltamatrix}								0 & 0 & 0 & 0 & 0 & 1 \\ \end{pmatrix}\,.
	\end{align}
 This breaking generates a large mass for both the tilde- and prime-quark sectors, leaving massless only the SM fermion components of the original fermionic fields:\footnote{Note that we take $\langle \Delta \rangle$
to be real.  The phases of the nonvanishing entries in 
Eq.~(\ref{Deltamatrix})
can all be made equal by an $SU(6)\times SU(3')$ transformation;
the remaining phase can be removed by a transformation under the $U(1)$
defined in Eq.~\eqref{BN'}.}
	\begin{align}
	\mathcal{L}\ni   \Lambda_{\rm{CUT}} \left\{ \kappa_q \, q_L' \tilde{q}_L
		+ \kappa_u \, u_L'^c \tilde{u}_L^c
		+ \kappa_d \, d_L'^c \tilde{d}_L^c \right\}+\, \text{h.c.}\,.
	\label{tildemass}
	\end{align}
Unless otherwise stated, we will assume in what follows that all $\kappa_i$ Yukawa couplings are $\mathcal{O} (1)$,   meaning all tilde and prime fermion masses are of order $\Lambda_{\rm{CUT}}$. Some tuning of the $\kappa_i $  values could be acceptable, though, as discussed further below.

The SM fermions get their masses  through  the usual SM Higgs doublet $\Phi$, which in this model is a singlet of $SU(6)\times SU(3')$,  \footnote{In this notation taken from unified models the contraction of the spinor indices is implicit, more precisely the first term would read $Q_L^T C \Phi U^c_L$, where $C=i\gamma_2\gamma_0$ is the charge conjugation matrix.}
\begin{align}
	\mathcal{L}\ni Y^{SM}_u \,Q_L \Phi U^c_L
		+ Y^{SM}_d \, Q_L \tilde{\Phi} D^c_L\, +\, \text{h.c.}\,,
		\label{SMYukawa}
	\end{align}
where $Y^{SM}_i$ denote the SM Yukawa couplings. Analogously,  the most general Lagrangian compatible with all symmetries discussed above allows 
us to write Yukawa couplings of the Higgs field to the prime-sector fermions, 
\begin{align}
	\mathcal{L}\ni  y'_u \,q'_L \Phi u'^c_L
		+ y'_d \, q'_L \tilde{\Phi} d'^c_L\, +\, \text{h.c.}\,.
		\label{primeYukawa}
	\end{align}
	 Eqs.~(\ref{SMYukawa}) and (\ref{primeYukawa}) induce  contributions to the tilde and fermion masses which are quantitatively irrelevant in comparison with those from  Eq.~(\ref{tildemass}). In addition, the couplings in Eq.~(\ref{primeYukawa}) will be absent for symmetry reasons in one of the models to be developed in this paper (model II in Sect.~\ref{model-II}).

Both $SU(3)_c$ and $SU(3)_{\rm{diag}}$ can now remain unbroken and confine at two different scales, $\Lambda_{\rm{QCD}}$ and $\Lambda_{\rm{diag}}$, with   $\Lambda_{\rm{diag}}\gg \Lambda_{\rm{QCD}}$. The task of achieving different values for the two confining scales and getting rid of the tilde sector or any other dangerous exotic sector is thus accomplished. 

	Note that the Yukawa-type Lagrangian in Eq.~(\ref{LDelta}) has an inherent global  $U(1)$ symmetry   under which only the prime fermions and the $\Delta$ field would transform -- a generalized Baryon number symmetry  in the prime sector, with charges
\beq
BN'\{\Delta,  q'_L, u'^c_L, d'^c_L\}=\{+1,+1,-1,-1\}\,.
\label{BN'}
\eeq
This symmetry is not chiral, thus not anomalous under $SU(3')$, and irrelevant to the strong CP problem. An associated pGB~\footnote{This symmetry is broken at loop level by $SU(2)_L$ sphalerons, in the same way that in the SM baryon number current is anomalous. For our purposes this effect is negligible.\label{BNN'sphalerons}}  results after spontaneous breaking, albeit with its couplings safely suppressed by the CUT scale and interacting only with the very heavy prime and tilde sectors.\footnote{As suggested in Ref.~\cite{Agrawal:2017ksf}, this type of pGB could be entirely removed by  gauging the U(1) group.  
There is no real need to implement this procedure in our case, though, given the strongly suppressed  couplings of this pGB.\label{foot_global_symm}}

 Finally, the theory below $\Lambda_{\text{CUT}}$ contains phenomenologically interesting bound states formed from the massless $\psi_{L,R}$ fermions, to be studied below. The spectrum of free eigenstates below the EW scale contains the usual SM spectrum, plus a harmless pGB and sterile neutrinos.

 \subsection*{$\theta'$ issue} 
The extension of the strong sector by the auxiliary external group $SU(3')$ brings a new $\theta'$ parameter into the game:
\beq
 \mathcal{L}\supset\theta_{6}\,\frac{\alpha_{6}}{8\pi}G_{6}\tilde{G}_{6} + \theta '\,\frac{\alpha '}{8\pi}G '\tilde{G '}\,\longrightarrow (\theta_{6}+\theta ')\,\frac{\alpha_{\rm{diag}}}{8\pi}G_{\rm{diag}}\tilde{G}_{\rm{diag}} + \theta_{6}\,\frac{\alpha_{c}}{8\pi}G_{c}\tilde{G}_{c}\,,
 \label{twotheta}
\eeq
where $G_i$ denote gauge field strengths with tensorial 
 indices omitted. $G_c$, $G_{6}$, $G ' $ and $G_{\rm{diag}}$ correspond respectively to 
the SM QCD gauge group,  $SU(6)$, $SU(3')$ and $SU(3)_{\rm{diag}}$.  
 While the rotation of the massless field $\Psi$ was designed to reabsorb $\theta_6$ and ultimately $\theta_c$, $\theta'$ may source back a SM strong CP problem through the contamination  to the visible sector via the $\Delta$ scalar.

	Indeed, at low energies the massless quark $\psi$ transforms as a $(3,\bar{3})$, therefore the phase $\theta_6$ cannot be fully reabsorbed in the Lagrangian since the chiral rotation that removes the $SU(3)_c$ $\theta$-term generates a new contribution to the $SU(3)_{\rm{diag}}$ topological term. 
	 Ref.~\cite{Gherghetta:2016fhp} acknowledges this issue (in the context of a different model which does not rely on massless fermions) and leaves it unsolved hoping that some UV completion solves it.  In what follows, we will determine and exhaustively analyze two UV solutions, via the simple addition of either
	\begin{itemize}
	 \item An extra massless fermion transforming only under $SU(3')$.
 \item A second bifundamental scalar field, which automatically endows PQ invariance to the above extension procedure.
  \end{itemize}
 The first solution is more in line with the spirit of the present paper, as {\it all} $\theta$ parameters inducing a strong CP problem  are made unphysical via massless fermions, and it is developed next.

 \subsection{Model I: Adding a massless fermion charged under $SU(3')$.}
 The $\theta'$ parameter of the auxiliary $SU(3')$ gauge group can be made unphysical by the addition of a massless fermion field $\chi$ that transforms as  a fundamental of $SU(3')$ and is an EW and $SU(6)$ singlet.  In other words, the field content for this solution is that previously shown in Tab.~\ref{model-matter-compact} plus the massless fermion $\chi$ with quantum numbers shown in Tab.~\ref{chi-model-matter-compact}.   Additional composite bound states will result from $\chi$, among them composite pseudoscalars with anomalous couplings\,---\,dynamical axions\,---\,whose masses are discussed further below.  
 
\begin{table}
\begin{align*}
\begin{array}{c||c|c|c|c||c|c|}
		& SU(6)		& SU(3)'	& SU(2)_L	& U(1)_Y & SU(3)_c	& SU(3)_\text{diag}	\\
\hline
\chi		& 1		& \Box		& 1		& 0 & 1 &  \Box	 		\\
\hline
\end{array}
\end{align*}
\caption{The table shows on the left (right) the quantum numbers above (under) the CUT scale for the massless $\chi$ quarks which absorbs $\theta'$ in model I.  }\label{chi-model-matter-compact}
\end{table}
 
 \subsubsection{Running of the coupling constants.}
\begin{figure}[t]
\centering
\includegraphics[scale=.7]{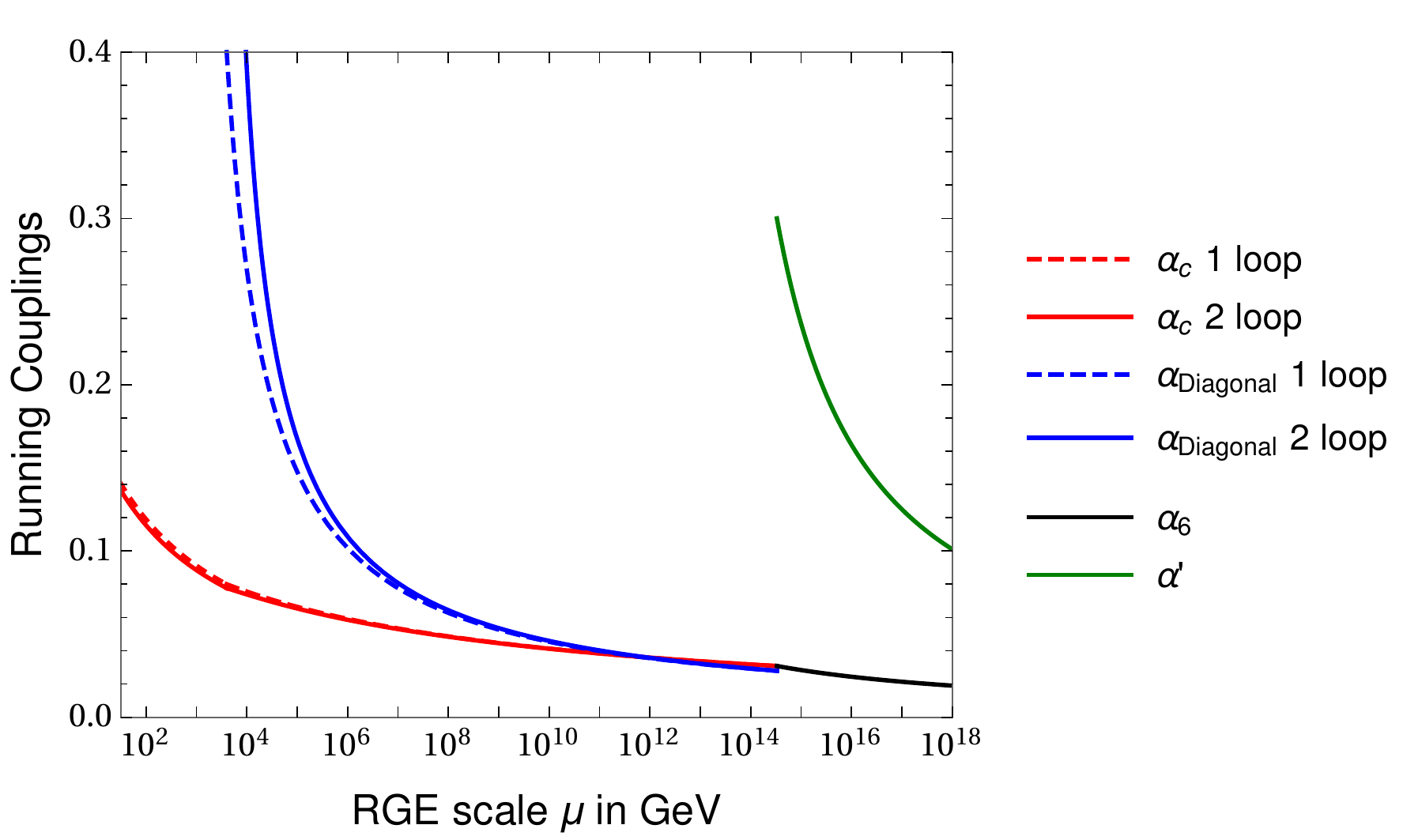}
\caption{Running of $\alpha_{\text{QCD}}$, $\alpha_{\rm{diag}}$, 
${\alpha_6}$, and $\alpha'$ in the model with the one extra $SU(3')$ fermion of Tab.~\ref{chi-model-matter-compact}. The full matter content is given by both Tabs.~\ref{model-matter-compact} and \ref{chi-model-matter-compact}.  The inputs used are  
$\alpha' (\Lambda_{\rm{CUT}}) = 0.3$ and $\Lambda_{\rm{CUT}}= 3.3\times10^{14}$ 
GeV for illustration, which results in $\Lambda_{\rm{diag}}= 4$ TeV (taken as a benchmark point). The solid (dashed) lines correspond to the two (one) loop results.
}
\label{Running-chi}
\end{figure}

The  CUT breaking pattern in Eq.~(\ref{SU6break}) imposes the following relations among the gauge couplings   
\begin{align}
	&\frac{1}{\alpha_{\rm{diag}}(\mu)}=\frac{1}{\alpha_6(\mu)}+\frac{1}{\alpha'(\mu)}\,, \qquad \text{at}\quad \mu=\Lambda_{\rm{CUT}}\,,
		\label{Coupling-relation}
	\end{align}
	with the constraint  
\begin{equation}
\alpha_c(\Lambda_{\rm{CUT}})= \alpha_6(\Lambda_{\rm{CUT}}) ,
\end{equation}	
where  $\alpha_c$, $\alpha_{\rm{diag}}$,  $\alpha'$ and $\alpha_6$ denote respectively the coupling strength of QCD, $SU(3)_{\rm{diag}}$, $SU(3')$ and $SU(6)$.
As shown in Fig.~\ref{Running-chi}, there is a discontinuity in the running of the coupling constants at the CUT-breaking scale  that allows $\alpha'$ to have large  values while reproducing the known QCD running at low scales. Those  $\alpha'$ values will seed a source of large axion masses, as discussed in Sect.~\ref{sec-SSI} further below. 

Although the relation in Eq.~(\ref{Coupling-relation}) imposes $\alpha_{\rm{diag}}(\Lambda_{\rm{CUT}})< \alpha_{c}(\Lambda_{\rm{CUT}})$,  the presence of  the SM $q_L$, $u_R$, and $d_R$ quarks at energies well below $\Lambda_{\rm{CUT}}$ slows down the running of QCD with respect to that of $SU(3)_{\rm{diag}}$. In this regime $\psi$ and $\chi$ are the only fields left charged under $SU(3)_{\rm{diag}}$ (the $\tilde{q}$ sector generically decouples as their mass scale is set by $ \Lambda_{\rm{CUT}}$, see Eq.~(\ref{tildemass})). As a consequence,  $\alpha_{\rm{diag}}$ runs faster and thus the $SU(3)_{\rm{diag}}$ group confines at a higher scale than $\Lambda_{\rm{QCD}}$, see Fig.~\ref{Running-chi}.  This mechanism easily achieves the separation of the two confining scales.  We computed both the one- and two-loop running and the latter actually reinforces the pattern, as illustrated in the figure   for the choice $\Lambda_{\rm{diag}}=4$ TeV. Lower values of $\Lambda_{\rm{diag}}$ are also phenomenologically acceptable, see Sect.~\ref{collider} below.

\subsubsection{Confinement of $SU(3)_{\rm{diag}}$ and pseudoscalar anomalous couplings to the confining interactions}
\begin{table}[h!]
\begin{align*}
\begin{array}{c|c|c}
	& SU(3)_c	& SU(3)_{\rm{diag}}	\\
\hline
\psi_{L}	& \Box		& \bar\Box	\\
\psi_{L}^c	& \bar\Box		& \Box	\\
\chi_{L}	& 1		& \Box	\\
\chi_{L}^c	& 1		& \bar\Box	\\
\end{array}
\end{align*}
\caption{The massless quark sector charged under $SU(3)_{\rm{diag}}$ remaining below the confining scale $\Lambda_{\rm{diag}}$.}
\label{massless-below-chi}
\end{table}
At the scale $\Lambda_{\rm{diag}}$, $SU(3)_{\rm{diag}}$ confines and the remaining massless fermions  ($\psi$ and $\chi$, see  Tab.~\ref{massless-below-chi})
 will form massive QCD-colored bound states.   
 In the limit in which $\alpha_c$ is switched off, the $SU(3)_{\rm{diag}}$ Lagrangian exhibits at the classical level a global flavor symmetry $U(4)_L \times U(4)_R\longrightarrow U(4)_V$.~\footnote{  $U(4)_V$ remains unbroken and contains as a subgroup the $SU(3)_c$ QCD gauge group.} The chiral symmetry is spontaneously broken by the quark condensates $\langle\bar\psi_{L}\psi_{R}\rangle$ and $\langle\bar\chi_{L}\chi_{R}\rangle$.
 This results in 16 (p)GBs,
 \begin{equation}
 16= 8_{c} + \bar 3_{c} + 3_{c} +1_c +1_c\,,
 \label{16decomp}
 \end{equation}
decomposed here in terms of their QCD charges.  There is a QCD octet plus a singlet  with flavor content $\bar{\psi}\psi$ $(\bar3_c\times3_c=8_c+1_c)$. The two QCD triplets, $3_c$ and $\bar{3}_c$, correspond to the combinations  $\bar{\psi}\chi$ and $\bar{\chi}\psi$. Finally, a color-singlet composite state is made out of $\bar{\chi}\chi$.
 The fourteen colored mesons in Eq.~(\ref{16decomp}) acquire large masses induced by gluon loops that are quadratically divergent and therefore sensitive to the cutoff scale $\Lambda_{\rm{diag}}$,\footnote{ They contribute to the running of the QCD coupling constant, but given their high masses their impact is unnoticeable. }  
\begin{align}
 m^2(8_c) &\approx \frac{9\,\alpha_\text{c}}{4\pi} \Lambda_\text{diag}^2\,, && m^2(\bar{3}_c ) \approx m^2(3_c ) \approx \frac{\alpha_\text{c}}{\pi} \Lambda_\text{diag}^2 \,.
 \label{colored-2}
\end{align} 
The remaining two QCD singlets will be denoted  here by $\eta'_\psi$ and $\eta'_{\chi}$ and are shown next to be dynamical axions. The  associated currents  are 
\begin{align}
j^\mu_{\psi_A}&=\bar{\psi}\gamup\gamma^5 t^9 \psi \equiv f_{\rm{d}}\,\partial^\mu\eta'_\psi\, , 
\qquad t^9=\frac{1}{\sqrt{6}}\mathds{1}_{3\times3}\,,\\
j^\mu_{\chi_A}&=\bar{\chi}\gamup\gamma^5\chi \equiv f_{\rm{d}}\,\partial^\mu\eta'_\chi\, ,
\end{align}
 where $f_{\rm{d}}$ denotes the $SU(3)_{\rm{diag}}$ pGB scale, with  $\Lambda_{\rm{diag}} \le 4\pi f_{\rm{d}} $. 
These classically conserved currents are broken at the quantum level by the $SU(6)$ and $SU(3')$  instantons, and so the currents are anomalous. The anomalous terms are
\begin{align}
\partial_\mu j_{\psi_A}^\mu & =-\sqrt{6}\frac{\alpha_{6}}{8\pi}G_{6}\tilde{G}_{6} \, \longrightarrow  -\sqrt{6}\,\frac{\alpha_{\rm{diag}}}{8\pi}G_{\rm{diag}}\tilde{G}_{\rm{diag}} -\sqrt{6}\,\frac{\alpha_{c}}{8\pi}G_{c}\tilde{G}_{c}\,,\\
\partial_\mu j_{\chi_A}^\mu & =\,-2\frac{\alpha '}{8\pi}G '\tilde{G '} \,\longrightarrow  -2\,\frac{\alpha_{\rm{diag}}}{8\pi}G_{\rm{diag}}\tilde{G}_{\rm{diag}}\,. 
\end{align}
 These anomalous terms modify the classical equations of motion of the  $\eta'_\psi$ and $\eta'_{\chi}$,
\begin{align}
f_{\rm{d}}\,\Box\,\eta'_\psi\, =-\sqrt{6}\frac{\alpha_{6}}{8\pi}G_{6}\tilde{G}_{6} \,,\\
 f_{\rm{d}}\,\Box\,\eta'_\chi\,=-2\frac{\alpha '}{8\pi}G '\tilde{G '} \,, 
\end{align}
and give rise to an effective Lagrangian,

\beq
\mathcal{L}\supset -\frac{\alpha_{6}}{8\pi}\,\frac{\sqrt{6}\,\eta'_\psi}{f_{\rm{d}}}\,G_{6}\tilde{G}_{6}-\frac{\alpha'}{8\pi}\,\frac{2\,\eta'_\chi}{f_{\rm{d}}}G'\tilde{G}'\,   \longrightarrow    
-\frac{\alpha_{c}}{8\pi}\,\frac{\sqrt{6}\,\eta'_\psi}{f_{\rm{d}}}\,G_{c}\tilde{G}_{c}-\frac{\alpha_{\rm{diag}}}{8\pi}\,\left(2\,\frac{\eta'_\chi}{f_{\rm{d}}}+\sqrt{6}\,\frac{\eta'_\psi}{f_{\rm{d}}}\right)G_{\rm{diag}}\tilde{G}_{\rm{diag}}\,.
\eeq
$\eta'_\psi$ and $\eta'_\chi$ are thus two dynamical axions.   It is to be stressed that the PQ scale in this model   is $f_{\text{d}} \sim \Lambda_{\rm{diag}}$ for both axions and not the much larger $\Lambda_{\rm{CUT}}$ scale. 
When the SM quarks are taken into account, the $\eta'_{\rm{QCD}}$ pseudoscalar meson is also present at energies below the QCD confinement scale, and the  effective Lagrangian of anomalous couplings reads

\beq
\mathcal{L}\supset    
-\frac{\alpha_{\rm{diag}}}{8\pi}\,\left(2\,\frac{\eta'_\chi}{f_{\rm{d}}}+\sqrt{6}\,\frac{\eta'_\psi}{f_{\rm{d}}}\right)G_{\rm{diag}}\tilde{G}_{\rm{diag}}\,- \frac{\alpha_{c}}{8\pi}\left(2\,\frac{\eta'_{\rm{QCD}}}{f_\pi}+\sqrt{6}\,\frac{\eta'_\psi}{f_{\rm{d}}}\right)G_{c}\tilde{G}_{c}\,,
\label{potential3}
\eeq
where    $ \Lambda_{\rm{QCD}}\le 4\pi f_\pi$. 
As a consequence, the two instanton-induced scales $\Lambda_{\rm{QCD}}$ and $\Lambda_{\rm{diag}}$ provide  a contribution to the masses of the pseudoscalars which have anomalous couplings; the corresponding effective potential is very well approximated by
\begin{align}
\mathcal{L}_{eff}=\Lambda_{\rm{diag}}^ 4\,\cos\left(2\,\frac{\eta'_\chi}{f_{\rm{d}}}+\sqrt{6}\,\frac{\eta'_\psi}{f_{\rm{d}}}\right)+\Lambda_{\rm{QCD}}^ 4\,\cos\left(2\,\frac{\eta'_{\rm{QCD}}}{f_\pi}+\sqrt{6}\,\frac{\eta'_\psi}{f_{\rm{d}}}\right)\, .
\label{axionpotential-2-first}
\end{align}
 It follows that there are only two sources of mass (disregarding corrections from SM quark masses) for three states coupling to anomalous currents: $\eta_{\rm{QCD}}'$, $\eta'_\psi$ and $\eta'_\chi$. In the absence of supplementary mass sources, one axion would  get a mass of order $\Lambda_{\rm{diag}}$ while another one would have remained almost massless, see Eqs.~(\ref{heavyaxion}) and Eqs.~(\ref{invisiblesaxion}), as often happens in models with dynamical axions.

\subsubsection{Impact of small-size instantons of the spontaneously broken CUT}
\label{sec-SSI}

 There is an additional  and putatively large contribution to the axion mass(es) in the presence of a spontaneously broken theory:  the small-size instantons (SSI) of the theory at the breaking scale, as pointed out long ago in Refs.~\cite{Holdom:1982ex, Dine:1986bg, Flynn:1987rs}  and very recently in  Ref.~\cite{Agrawal:2017ksf}. SSI can induce a large mass  even for perturbative theories if the breaking scale is large enough to overcome the exponential suppression of instanton effects. 
 In our model, the instantons of the color-unified theory in Eq.~(\ref{SU6break}) near the $\Lambda_{\rm{CUT}}$ scale provide automatically this third source of  axion mass.  The $SU(6)$ SSI can be neglected because of the 
 smallness of $\alpha_6$ at $\Lambda_{\rm{CUT}}$ (e.g. see Fig.~\ref{Running-chi}) and the analysis below will focus on the $SU(3')$ SSI instantons. 
 
  It is well known~\cite{tHooft:1976snw, Callan:1977gz, Shifman:1979uw,Shifman:1994ee} that, in the absence of fermions,
  the effective Lagrangian that describes instanton configurations for a pure Yang-Mills theory $SU(N_c)$ induces a scale  $\Lambda_{\rm{inst}}$ in the instanton potential  given by
 \beq
\Lambda_{\rm{inst}}^4=  \int \frac{d\rho}{\rho^5}\, D[\alpha'(1/\rho)]\,, \label{InstantonCallan}
 \eeq
 where $\rho$ is the instanton size, $D[\alpha']$ is the dimensionless instanton density,

 \beq
D[\alpha'(1/\rho)]= C_{inst}\left( \frac{2\pi}{\alpha'(1/\rho)}\right)^{2N_c} e^{-\nicefrac{2\pi}{\alpha'(1/\rho)}}\,.
 \eeq
  The constant $C_{inst}$  reads \cite{Bernard:1979qt,Sannino:2018suq} 
  \beq
  C_{inst}(N_c)=\frac{4}{\pi^{2}}\frac{ 2^{- 2 N_{c}} e^{- c(1) - 2 \left(N_{c} - 2\right) c(\nicefrac{1}{2})}}{\left(N_{c} - 2\right)! \left(N_{c} - 1\right)!}\,,
  \eeq
  and the function $c(x)$  defined in Ref.~\cite{tHooft:1976snw} such that $c(\nicefrac{1}{2}) = 0.145873$ and $c(1) = 0.443307$.   
  For the  $SU(3')$ instantons of our model  $C_{inst}=0.0015$.\footnote{This value differs from that used in Ref.~\cite{Agrawal:2017ksf} ($C_{inst}=0.1$) that was taken directly from  the original 't Hooft's computation in Ref.~\cite{tHooft:1976snw}, for which it was later shown~\cite{Bernard:1979qt} that  the factor $2^{- 2 N_{c}}$ was missing.  See also Erratum in Ref.~\cite{tHooft:1976snw}. }  In order to compute the integral in Eq.~(\ref{InstantonCallan}),   the running of the coupling constant $\alpha'(\mu)$ must be included. At one loop this reads 
  \beq
  \frac{2\pi}{\alpha'(\mu)}=b\,\ln\left(\mu/\Lambda_{\rm{CUT}}\right)+\frac{2\pi}{\alpha'_{\rm{CUT}}}\,,
  \eeq
  where 
  $\alpha_{\rm{CUT}}'\equiv\alpha'(\Lambda_{\rm{CUT}})$ and $b$  is the one-loop $\beta$-function coefficient.
    For the spontaneously broken theory, only the SSI instantons  with size $\le 1/\Lambda_{\rm{CUT}}$ are relevant,  
   \beq
\Lambda_{\rm{SSI}}^4=  C_{inst}e^{-\nicefrac{2\pi}{\alpha'_{\rm{CUT}}}}\, \int_0^{1/\Lambda_{\rm{CUT}}} \frac{d\rho}{\rho^5}\, (\rho\Lambda_{\rm{CUT}})^b \left(-b\,\ln\left(\rho\Lambda_{\rm{CUT}}\right)+\frac{2\pi}{\alpha'(\Lambda_{\rm{CUT}})} \right)^6 \,. \label{Instanton1loopNoFerm}
 \eeq
  This has the form
\beq
\Lambda_{\rm{SSI}}^4 = C_{inst}\,f(\alpha'_{\rm{CUT}},b)\,e^{-\frac{2 \pi }{\alpha'_{\rm{CUT}}}}  \Lambda_{\rm{CUT}}^4\,, \label{YangResult}
 \eeq
 where  $f(\alpha',b)$ is given by 
\begin{multline}
  f(\alpha',b)= \frac{16}{\alpha'^6 (b-4)^7} \left(45 \alpha'^6 b^6+90 \pi  \alpha'^5 b^5 (b-4)+90 \pi ^2
   \alpha'^4 b^4 (b-4)^2 \right.\\
    \left.+60 \pi ^3 \alpha'^3 b^3 (b-4)^3+30 \pi ^4 \alpha'^2 b^2 (b-4)^4+12 \pi ^5 \alpha' b
   (b-4)^5+4 \pi ^6 (b-4)^6\right)\,.\label{ffunction}
\end{multline}
For instance for the benchmark value $\alpha'_{\rm{CUT}}= 0.3$, the value of $SU(3')$ SSI-induced scale in the absence of fermions (for which $b=10$)  is
\beq
\Lambda_{\rm{SSI}}^4= 3.0\times 10^{-5}\,\Lambda_{\rm{CUT}}^4\,.
\label{Leff-fermionless}
\eeq

\begin{figure}[t]
\centering
\begin{centering}
\hfill \includegraphics[width=.43\textwidth]{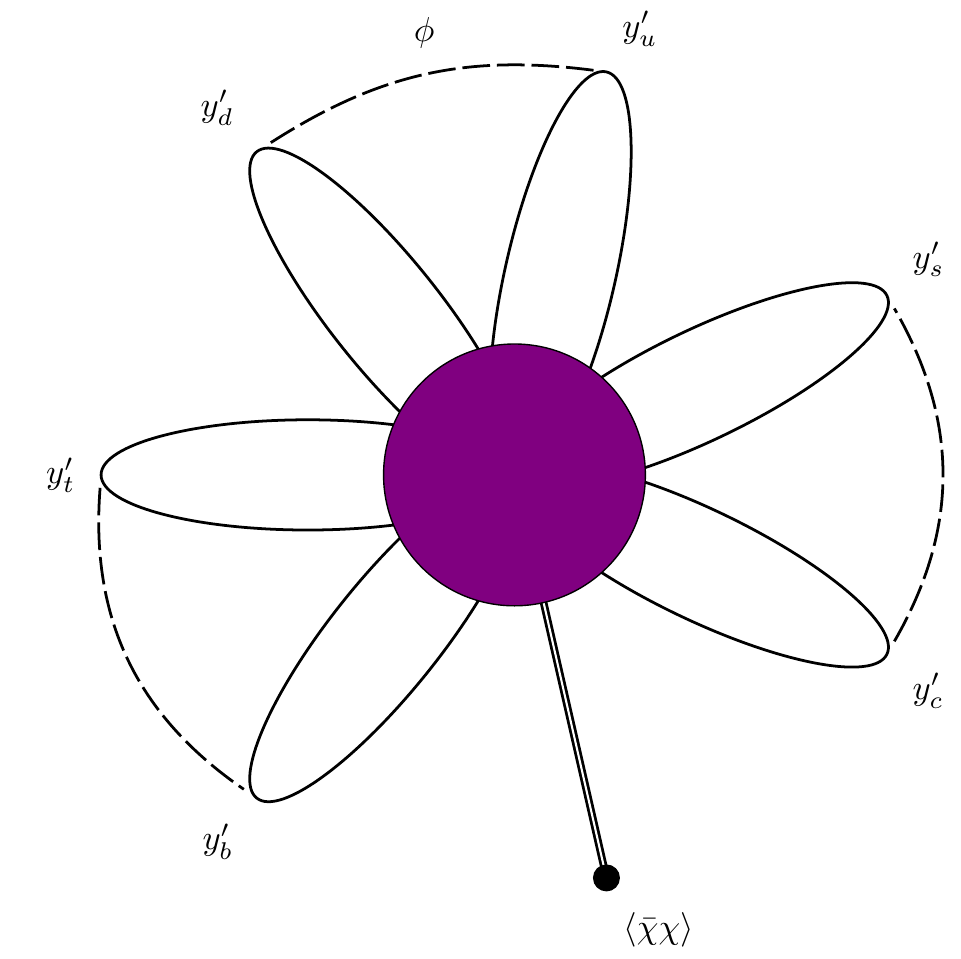} \hfill \includegraphics[width=.47\textwidth]{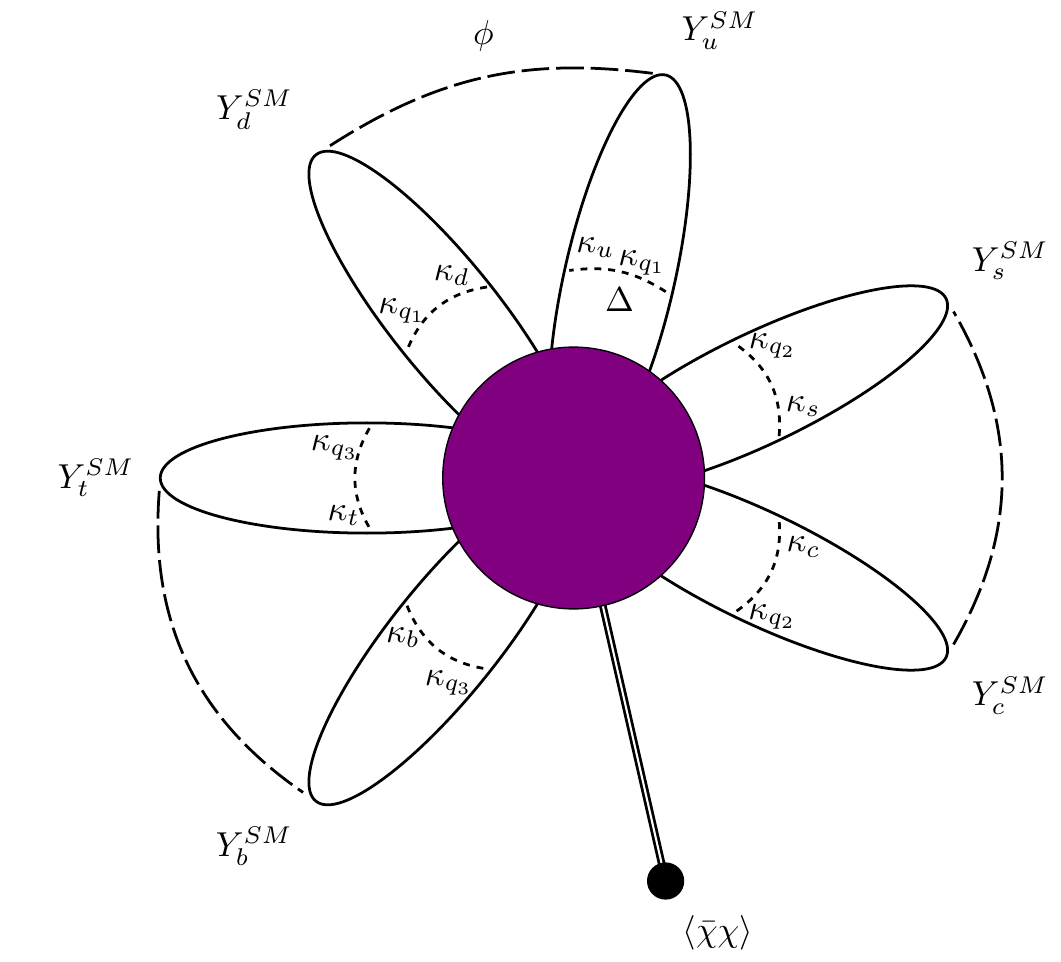}\hfill\hfill
\caption{Instanton contributions in model I. The long dashed lines connecting the $y'_i \,\bar{q}'_L \Phi  \,u_{L}^{\prime c}$ 
 interactions correspond to $\phi$ propagators, while the short dashed lines depict propagators of the $\Delta$ scalar. } \label{instantons_model1}
\end{centering}
\end{figure}

Nevertheless,  the presence of fermions dramatically changes  the value of this scale~\cite{Shifman:1979uw}. A suppression factor  appears, 
 which results from the interplay of the instantons of the theory and the fermionic spectrum. For the $SU(3')$ theory under consideration, the prime-fermion Yukawa couplings  are relevant. For generic values of all Yukawa couplings of order one, the dominant contribution stems from the $y_i'$ couplings in Eq.~(\ref{primeYukawa}). They are illustrated by the one-instanton ``flower" contribution in Fig.~\ref{instantons_model1}a. Its impact is to suppress the pure gauge result in Eq.~(\ref{InstantonCallan}) by two factors: the $\chi$ chiral condensate  and the $y'_i$ Yukawas couplings of the primed sector,
 	\beq
\Lambda_{\rm{SSI}}^4 =  -\int \frac{d\rho}{\rho^5}\, D[\alpha'(1/\rho)]  \left(  \frac{2}{3}\pi^ 2\rho^ 3\, \langle\bar \chi \chi\rangle \right) \frac{1}{(4\pi)^6}\prod_i  y'^{\,i}_u y'^{\,i}_d\,. \label{InstantonShifman-chi0}
 \eeq
$\langle\bar \chi \chi\rangle$ is the order parameter controlling $SU(3)_{\rm{diag}}$ chiral symmetry breaking and thus expected to be 
$\langle\bar \chi \chi\rangle\simeq-\Lambda_{\rm{diag}}^3$.  For $\rho\le 1/\Lambda_{\rm{CUT}}$,  the product $\rho^ 3\, \langle\bar \chi \chi\rangle \ll 1$ reduces the SSI-induced scale by orders of magnitude,  with 
\beq
\Lambda_{\rm{SSI}}^4=  \frac{2\pi^2}{3}\,C_{inst}\,\Lambda_{\rm{CUT}}^{b}\,\Lambda_{d}^3\, e^{-\nicefrac{2\pi}{\alpha_{\rm{CUT}}'}}\,\frac{1}{(4\pi)^6}\prod_i  y'^{\,i}_u y'^{\,i}_d\,\int_0^{1/\Lambda_{\rm{CUT}}} d\rho\, \rho^{b-2} \left(-b\,\ln\left(\rho\Lambda_{\rm{CUT}}\right)+\frac{2\pi}{\alpha_{\rm{CUT}}'} \right)^6\,,
 \label{InstantonShifman-chi}
 \eeq 
 where, in the presence of  $N_f$ Dirac fermions, 
 \beq
C_{inst}(N_f,N_c)=\frac{4 \cdot 2^{- 2 N_{c}} e^{c_{1/2} (- 2 N_{c} + 2 N_{f})}}{\pi^{2} \left(N_{c} - 2\right)! (N_{c} - 1)!} e^{- c_1 + 4 c_{1/2}}\,.
\eeq
 The integral in Eq.~(\ref{InstantonShifman-chi}) can be computed exactly, although a good estimation follows from the approximation 
   \beq
 \left(1+\frac{-b\,\alpha_{\rm{CUT}}'}{2\pi}\,\ln\left(\rho\Lambda_{\rm{CUT}}\right) \right)^6 \simeq1+ 6\,\frac{-b\,\alpha_{\rm{CUT}}'}{2\pi}\,\ln\left(\rho\Lambda_{\rm{CUT}}\right)\,,  \label{Approx}
 \eeq
 which leads to 
 \begin{equation}
 \Lambda_{\rm{SSI}}^4\simeq \frac{\pi}{96}\,C_{inst}\,\Lambda_{\rm{diag}}^3\,\Lambda_{\rm{CUT}}\,e^{-\nicefrac{2\pi}{\alpha_{\rm{CUT}}'}}\,\frac{3b\,\alpha_{\rm{CUT}}'+(b-1)\pi}{\alpha_{\rm{CUT}}'^{\,6}(b-1)^2}\prod_i  y'_{\,u_i} y'_{\,d_i}\,.
 \end{equation}
  For the benchmark $\alpha'_{\rm{CUT}}=0.3
 $, and substituting $N_f=7$ and $b=16/3$, this gives 
 \beq
 \Lambda_{\rm{SSI}}^4\simeq 4.5\times 10^{-10}\,\Lambda_{\rm{diag}}^3\,\Lambda_{\rm{CUT}}\,. 
\eeq 
The complete computation including fermions can be compared with the one in in Eq.~(\ref{Leff-fermionless}),  a strong suppression by a factor of order $(\Lambda_{\rm{diag}}/\Lambda_{\rm{CUT}})^3$ is in fact present.  
 
 There is in addition a subdominant contribution to the SSI scale,  suppressed by the $\chi$ chiral condensate, the product of $\kappa_i$ Yukawa coupling of the prime-fermion sector in Eq.~(\ref{LDelta}) {\it and}  the product of SM Yukawa couplings. This contribution is illustrated by the instanton ``double flower'' in Fig.~\ref{instantons_model1}b,  and given by
     	\beq
\delta \Lambda_{\rm{SSI}}^4 =  -\int \frac{d\rho}{\rho^5}\, D[\alpha'(1/\rho)]  \left(  \frac{2}{3}\pi^ 2\rho^ 3\, \langle\bar \chi \chi\rangle \right) \frac{1}{(4\pi)^{18}}\prod_i  Y^{SM}_{\,u_i} Y^{SM}_{\,d_i}\,(\kappa_q^i)^2 \kappa_u^i \kappa_d^i \,, 
\label{InstantonShifman-Delta2}
 \eeq
where the power of the $4\pi$ factor results from the 6 SM Yukawa couplings and the 12 $\kappa_i'$ couplings in the product.
 For $\rho\le 1/\Lambda_{\rm{CUT}}$, this contribution   is well approximated by 
 \beq
\delta \Lambda_{\rm{SSI}}^4=   \frac{2}{3}\,\frac{\pi^ 2}{(4\pi)^{18}}C_{inst}\Lambda_{\rm{CUT}}^{b}e^{-\frac{2\pi}{\alpha'_{\rm{CUT}}}}\prod_i  Y^{SM}_{\,u_i}Y^{SM}_{\,d_i} \,\kappa_q^{i\,^2} \kappa_u^i \kappa_d^i  \int_0^{1/\Lambda_{\rm{CUT}}} d\rho\, \rho^{b-2} \left(-b\,\ln\left(\rho\Lambda_{\rm{CUT}}\right)+\frac{2\pi}{\alpha'(\Lambda_{\rm{CUT}})} \right)^6 \,.
 \label{Instanton1loop2S}
 \eeq
 In summary,  putting together the dominant and subdominant instanton contributions discussed, the  SSI scale is given by
    	\beq
 \Lambda_{\rm{SSI}}^4 =  -\int \frac{d\rho}{\rho^5}\, D[\alpha'(1/\rho)]  \left(  \frac{2}{3}\pi^ 2\rho^ 3\, \langle\bar \chi \chi\rangle \right) \left\{\frac{1}{(4\pi)^{6}}\prod_i  y'^{\,i}_u y'^{\,i}_d\,\,+\frac{1}{(4\pi)^{18}}\,\prod_i  Y^{SM}_{\,u_i} Y^{SM}_{\,d_i}\,(\kappa_q^i)^2 \kappa_u^i \kappa_d^i \right\}\,. 
\label{InstantonShifman-chi-total}
 \eeq

 Overall, the size of the new scale $\Lambda_{\rm{SSI}}$ is  quite sensitive to the value of the $SU(3')$
  coupling constant at the CUT-breaking scale.  Fig.~(\ref{LambdaSSI}) illustrates the $\eta_\chi'$ axion mass induced by the small size instantons.   
   For the benchmark examples studied, $\Lambda_{\rm{SSI}}$ significantly affects the properties of the pseudoscalars. 
   It provides a new contribution to the  effective potential of the form  \begin{equation}
 \delta \mathcal{L}_{eff}=\Lambda_{\rm{SSI}}^ 4\,\cos\left(2\,\frac{\eta'_\chi}{f_{\rm{d}}}\right)\,.
 \label{SSI-chi}
\end{equation}
A mass is thus generated for the $\eta'_\chi$ axion, given by
\begin{equation}
m_{\eta'_{\chi}}^2 \sim 2.6 \times10^{-7}\,\, \Lambda_{\rm{diag}}\, \Lambda_{\rm{CUT}}\,,
\label{meta'-SSI-chi}
\end{equation}
where the replacement $\Lambda_{\rm{diag}}\simeq 4 \pi f_{\rm{d}}$ has been used.
 The other dynamical axion of the theory has been shown to acquire a mass of order $\Lambda_{\rm{diag}}$, see Eq.~(\ref{axionpotential-2}). Both dynamical axions have  thus acquired masses of order TeV, as a direct and unavoidable consequence of the instanton potentials inherent to the theory. 
 \vspace{-0.2cm}
 \subsubsection*{How light can the axion that couples to SSI become?}
 \vspace{-0.3cm}
 \noindent The $y_i'$ values are very relevant for the size of SSI scale and a priori provide the dominant contribution, as explained above. Nevertheless, 
 the mass spectrum and thus the running of coupling constants is basically unaffected by them. Should the $y_i'$ couplings be negligible,  $\Lambda_{\rm{SSI}}$ would be determined by the second term in Eq.~(\ref{InstantonShifman-chi-total}). How small can this scale become, and thus how light can the axion coupled to it be? For vanishing $y_i'$ values and generic $\kappa_i$ couplings of $\mathcal{O}(1)$, the product of  $Y^{SM}_i/4\pi$ factors in the second term in Eq.~(\ref{InstantonShifman-chi-total}) would  suppress the $\eta'_\chi$ axion mass values illustrated in Fig.~\ref{LambdaSSI} by a factor of $\mathcal{O}(10^{-6})$. In other words, the $\eta'_\chi$ axion would then have a mass smaller than  $m_{\eta'_\psi}\sim \Lambda_{\rm{diag}}$ and under the TeV scale.  Can this axion be lighter still by assuming both negligible $y_i'$ couplings and small $\kappa_i$ couplings? The answer is yes although this possibility is limited by the fact that the $\kappa_i$ couplings  determine essentially the spectrum of the tilde sector which, if lighter, may strongly impact the  running of the coupling constants. To get an estimate neglecting $y_i'$ couplings,  with all $\kappa_i$ values of $\mathcal{O}(10^{-1})$ there is little impact on the running while the $\Lambda_{\rm{SSI}}$ scale is low enough  to make the  $\eta'_\chi$ axion accessible at colliders;  $\kappa_i$ values of $\mathcal{O}(10^{-2})$ also  allow a realistic setup and would bring down the $\eta'_\chi$ mass even under the GeV regime and maybe it could be as light as an invisible axion. Other patterns of $\kappa_i$  couplings may also be possible.   Nevertheless,  in what follows we will not pursue this {\it ad hoc} avenue of  fine-tuning the $y_i'$ and $\kappa_i$ couplings to very small values. Unless stated otherwise, $\mathcal{O}(1)$ values  will be assumed for all prime-fermion Yukawa couplings.

\subsubsection{Solution to the strong CP problem.}
It is pertinent to briefly re-check the status of the strong CP problem  after taking into account the impact of  the SSI of the spontaneously 
broken symmetry discussed above.  
 Any new mass term for the axions breaks the PQ symmetry and therefore perturbs the axion potential; it is then important to verify that the vevs of the axions remain in the CP-conserving minimum, solving the strong CP problem.
 Indeed, this is the case with our color-unified proposal as, according to Eq.~(\ref{twotheta}), the potential including $\theta_i$ dependencies explicitly reads
\begin{align}
\mathcal{L}_{eff}=\Lambda_{\rm{SSI}}^ 4\,\cos\left(-2\,\frac{\eta'_\chi}{f_{\rm{d}}}+\bar\theta'\right)\,+\Lambda_{\rm{diag}}^ 4\,\cos\left(-2\,\frac{\eta'_\chi}{f_{\rm{d}}}-\sqrt{6}\,\frac{\eta'_\psi}{f_{\rm{d}}}+\bar\theta'+\bar\theta_6\right)+\Lambda_{\rm{QCD}}^ 4\,\cos\left(-\sqrt{6}\,\frac{\eta'_\psi}{f_{\rm{d}}}+\bar\theta_6\right)\,. 
\label{finalpotential1}
\end{align}
For this potential, the minimum is CP-conserving: 
 \begin{align}
\left\langle\bar{\theta'}-\,2\,\frac{\eta'_\chi}{f_{\rm{d}}}
\right\rangle=0 \,,
	&& 
\left\langle\bar{\theta}_{6}-\,\sqrt{6}\,\frac{\eta'_\psi}{f_{\rm{d}}}
\right\rangle=0\,,
\label{thetaminima-2}
\end{align} 
since all $\theta_i$ dependences cancel.  A word of caution is pertinent as  the exact dependence of the potential on the phases of the different couplings in the Lagrangian which participate in fermion mass generation ($\kappa_i,\, Y^{SM}_i, \,y'_i,\,\dots$) remains to be computed. Nevertheless, the two massless fermions $\Psi$ and $\chi$ guarantee that at energies above CUT the two parameters $\bar{\theta}_{6}$ and $\bar{\theta'}$ are unphysical. Below CUT, the spectrum of the theory is exclusivley the SM one plus massless fermions, and the EW SM contributions are known to be negligible~\cite{Ellis:1978hq}, even if a mismatch remained in spite of the low-energy presence of the massless quarks. An explicit computation of the threshold effects is elaborate and it is left for future work.  Note that without the presence of the second PQ mechanism, that is, without the presence of the $\eta'_\chi$ field and its vev, all $\theta_i$ in Eq.~(\ref{twotheta}) would not have been reabsorbed, while Eq.~(\ref{thetaminima-2}) demonstrates that its inclusion does ensure a CP-conserving minimum.

It is very positive that in this model there is no contribution to the EW hierarchy problem coming from axion physics. No potential connects the EW and axion scales:  the PQ scale $f_a$
\footnote{
The PQ scale (usually denoted by $f_{PQ}$) and $f_a$ differ by a model-dependent factor stemming from the relative strength of the axion coupling to gluons. Here we disregard the distinction between $f_a$ and $f_{PQ}$.
}
 is set by $\Lambda_{\rm{diag}}$  and not $\Lambda_{\rm{CUT}}$,  and  all axions are dynamically generated.  
 This is a feature that our model I shares with the original axicolor model, and in general with models of composite dynamical axion(s). There remains instead the customary fine-tuning  in spontaneously broken unified theories, as $\Lambda_{\rm{CUT}}$ and the EW scale are connected via the scalar potential, but the latter does not communicate to our PQ mechanism.

\subsubsection{Computation of the pseudoscalar mass matrix: $\eta'_\chi$,  $\eta'_\psi$, $\eta'_{\rm{QCD}}$ and light spectrum}
\label{3mesons}
After the replacement of the pGBs with anomalous couplings by their physical excitations, $\eta'_\chi\to \langle \eta'_\chi \rangle + \eta'_\chi $, $\eta'_\psi \to \langle\eta'_\psi\rangle +\eta'_\psi $, the effective low-energy Lagrangian for the axions and the SM $\eta'_{\rm{QCD}}$ field is given by 
(disregarding the effects of SM quark masses)
\begin{figure}[h!]
\centering
\includegraphics[scale=.55]{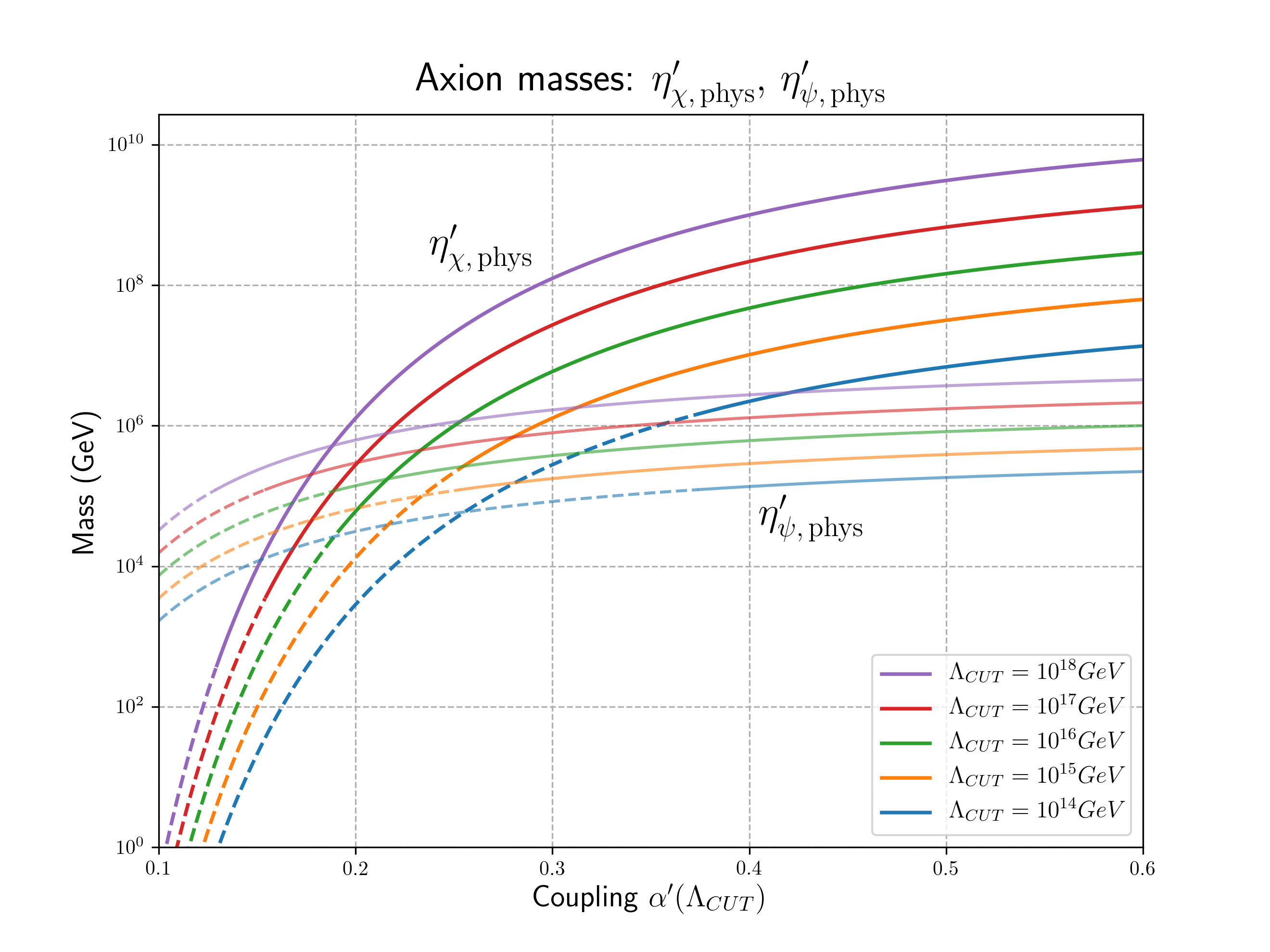}
\caption{Model with massless $\chi$. The $\eta'_\chi$ axion mass induced by the $SU(3')$ small size instantons at the scale $\Lambda_{\rm{CUT}}$ is shown together with the smaller mass for the dynamical $\eta'_\psi$ axion sourced by $\Lambda_{\rm{diag}}$ instantons. The $\alpha_{\rm{CUT}}'$ values corresponding to solid lines are allowed, while those for dashed lines correspond to excluded values, for which $SU(3)_{\rm{diag}}$ would confine below  $2.9$ TeV, ruled out by searches for scalar octets at LHC~\cite{Aaboud:2017nmi} (see Sec.~\ref{collider}). }
\label{LambdaSSI}
\end{figure}

\begin{align}
\mathcal{L}_{eff}=\Lambda_{\rm{SSI}}^ 4\,\cos\left(2\,\frac{\eta'_\chi}{f_{\rm{d}}}\right)\,+\Lambda_{\rm{diag}}^ 4\,\cos\left(2\,\frac{\eta'_\chi}{f_{\rm{d}}}+\sqrt{6}\,\frac{\eta'_\psi}{f_{\rm{d}}}\right)+\Lambda_{\rm{QCD}}^ 4\,\cos\left(2\,\frac{\eta'_{\rm{QCD}}}{f_\pi}+\sqrt{6}\,\frac{\eta'_\psi}{f_{\rm{d}}}\right)\, .
\label{axionpotential-2}
\end{align}
Expanding to second order in the fields yields the following mass matrix:
\begin{align}
M^2_{\eta'_\chi,\,\eta'_\psi,\,\eta'_{\rm{QCD}}}=\,\left(\begin{array}{ccc}
4\frac{\left(\Lambda_{\rm{SSI}}^ 4+\Lambda_{d}^ 4\right)}{f_{\rm{d}}^2}&2\sqrt{6}\,\frac{\Lambda_{d}^ 4}{f^2_d}&0\\ 
2\sqrt{6}\,\frac{\Lambda_{d}^ 4}{f^2_d} & 6\,\frac{\left(\Lambda_{d}^ 4+\Lambda_{\rm{QCD}}^ 4\right)}{f_{\rm{d}}^2} & 2\sqrt{6}\,\frac{\Lambda_{\rm{QCD}}^ 4}{f_{\pi}f_{\rm{d}}}\\ 
0&2\sqrt{6}\,\frac{\Lambda_{\rm{QCD}}^ 4}{f_{\pi}f_{\rm{d}}}& 4\frac{\Lambda_{\rm{QCD}}^ 4}{f_{\pi}^2}
\end{array}\right)\,.
\label{massmat}
\end{align}
Assuming the hierarchy of scales $\frac{\Lambda_{\rm{SSI}}^ 4}{f_{\rm{d}}^2}\gg\frac{\Lambda_{d}^ 4}{f_{\rm{d}}^2}\gg\frac{\Lambda_{\rm{QCD}}^ 4}{f_\pi^2}$ (valid for certain regions of parameter space, see Fig.~\ref{LambdaSSI}), the  three mass eigenvalues are 
\begin{align}
m^2_{\eta'_{\chi,phys}}\simeq4\frac{\Lambda_{\rm{SSI}}^ 4}{f_{\rm{d}}^2}\,,\qquad
m^2_{\eta'_{\psi,\,phys}}\simeq6\frac{\Lambda_{d}^ 4}{f_{\rm{d}}^2}\,,\qquad
m^2_{\eta'_{QCD\,phys}}\simeq4\frac{\Lambda_{\rm{QCD}}^ 4}{f_\pi^2}\,.
\label{pseudoscalarmasses-I}
\end{align}
As advertised, only the usual QCD $\eta'_{QCD\,phys}$ remains as a light eigenstate, while the two composite dynamical axions will be very heavy: one will have typically a mass of tens of TeV and the other will be orders of magnitude heavier.\footnote{As previously discussed, the heavier axion could be made much lighter by fine-tuning the Yukawa couplings for the prime-sector fermions, see Sect.~\ref{sec-SSI}.}

\subsubsection*{Low energy spectrum and observable effects}

The two dynamical axions ${\eta'_{\psi,\,phys}}$ and $\eta'_{\chi,phys}$ are typically heavier than the TeV scale and thus not easy to directly observe at the Large Hadron Collider (LHC).       Under a few TeV the spectrum of the theory contains:
\begin{itemize}
\item The SM pseudoscalar meson $\eta'_{QCD\,phys}$, plus the rest of the SM hadronic spectrum.
\item   The exotic QCD-colored ``pions''\,---\,color octets and color triplets\,---\,whose masses are given in Eq.~(\ref{colored-2}) as $m^2\sim \alpha_c \Lambda^2_\text{diag}$. With masses naturally lighter than the TeV scale, these QCD-colored pions can be easily produced at the LHC. 

 \item The two sterile fermions stemming from the 20- representation 
$\Psi$. They are basically invisible as their interactions with the visible world are suppressed by $\Lambda_{\rm{CUT}}$, which is much larger than $\Lambda_{\rm{diag}}$ without any tunning.
 \item Possibly, a GB associated with generalized baryon number. This GB is harmless as its interactions are suppressed by $\Lambda_{\rm{CUT}}$. It can also easily be made arbitrarily heavy by gauging that global symmetry.

\end{itemize} 

A further pertinent comment is that the massless $\chi$ fermions  may themselves acquire an effective mass due to the instantons of  $SU(3)_{\rm{d}}$ and  to $SU(3')$ SSI instantons, similar to the effective mass in QCD for a hypothetically massless SM up quark. 
 This is an interesting question which will not be further developed in this paper.

The very interesting phenomenological bounds and detection prospects for the exotic QCD-colored mesons will be quite similar to those applying to the next model.  The ensemble will then be briefly developed in  Sect.~\ref{pheno} further below. The same applies to the cosmological consequences of the two color-unified UV completions developed in this paper.

 \subsection{Model II: Addition of a second $\Delta$ scalar.}
 \label{model-II}

This solution to the $\theta'$ problem is an alternative to extending the spectrum by a massless fermion, discussed in the previous subsection. In this second model no extra fermion is added to the $SU(6)\times SU(3')$ Lagrangian, while a second $\Delta$ field  will be considered instead. The spectrum is that in Tab.~\ref{model-matter-compact} albeit with the scalar line duplicated,   $\Delta \rightarrow \{ \Delta_1, \, \Delta_2\}$. This simple extension allows the implementation of a  PQ symmetry which reabsorbs the $\theta'$ contribution to the strong CP problem. The corresponding PQ symmetry is automatic  if  the terms in Eq.~(\ref{primeYukawa}) are omitted and  Eq.~(\ref{LDelta}) is replaced by 
	\begin{align}
	\mathcal{L}\ni \kappa_q \,q_L' \Delta_1^* Q_L
		+ \kappa_u \, u_L'^c \Delta_2\, U_L^c 
		+  \kappa_d \, d_L'^c \Delta_2 D_L^c \, + \text{h.c.}\,.
		\label{LDelta12}
	\end{align}
 This Lagrangian is invariant under two independent abelian global symmetries; one of them is anomalous with respect to $SU(3')$ and corresponds to the PQ charge assignment~\footnote{ The Lagrangian possesses another $U(1)$ symmetry, namely the generalized baryon number symmetry defined in Eq.~\eqref{BN'}, which is however non-anomalous under $SU(3')$. See  footnotes~\ref{BNN'sphalerons} and \ref{foot_global_symm} on the harmless consequences of the non-anomalous global symmetry.}
\beq
PQ\{\Delta_1, \Delta_2, q'_L, u_L'^c, d_L'^c\}=\{+1,-1,+1,+1,+1\}\,.
\label{charges-Delta}
\eeq
The vevs
of $\Delta_1$ and $\Delta_2$ generalize the CUT spontaneous breaking in Eq.~(\ref{Deltamatrix}) and  at the same time break spontaneously the PQ symmetry; therefore, this PQ scale coincides with the CUT scale. This distinguishes model 2 from model 1, as in the latter the PQ scale coincided with $\Lambda_{\text{diag}}$. A pGB\,---an elementary axion---\,is generated at this stage. 
The corresponding PQ conserved current is given by 
\beq
j_{PQ}^\mu = \left[\overline{q'_L}\gamup q'_L+\overline{u_L'^c}\gamup u_L'^c+\overline{d_L'^c}\gamup d_L'^c +i(\Delta_1\dmup \Delta_1^* - \Delta_2\dmup \Delta_2^* - \hc )\right]\,. 
\eeq
The $\Delta_i$ fields are parameterized as
\beq
\Delta_1\equiv \frac{1}{\sqrt{2}}(\rho_1+v_{\Delta_1})e^{ia_1/v_{\Delta_1}}\,,\qquad \Delta_2\equiv \frac{1}{\sqrt{2}}(\rho_2+v_{\Delta_2})e^{ia_2/v_{\Delta_2}}\,,
\eeq
where $v_{\Delta_1}$ and $v_{\Delta_2}$ denote respectively the $\Delta_1$ and $\Delta_2$ vevs, both of which we take to be real
for simplicity.
 Decoupling the heavy radial modes, the PQ current reads
\beq
j_{PQ}^\mu \supset v_{\Delta_1}\partial^\mu a_1-v_{\Delta_2}\partial^\mu a_2\equiv f_{a}\,\partial^\mu a\,,
\eeq
where the elementary axion field $a(x)$ corresponds to the GB combination 
\begin{equation}
a(x) = \frac{1}{f_a}(v_{\Delta_1}a_1(x) -v_{\Delta_2}a_2(x))\,,
\end{equation}
with
\begin{equation}
f_{a}=\Lambda_{\rm{CUT}}=\sqrt{v_{\Delta_1}^2+v_{\Delta_2}^2}\,.
\label{fPQ}
\end{equation}
This classically exact PQ symmetry  
is broken at the quantum level by the $SU(3')$ anomaly, which at lower energies translates into an  anomalous current for the  $SU(3)_{\rm{diag}}$ gauge theory.
\beq
\partial_\mu j_{PQ}^\mu=\frac{\alpha'}{8\pi}N'G'\tilde{G}' \longrightarrow \frac{\alpha_{\rm{diag}}}{8\pi}N_{\rm{diag}}G_{\rm{diag}}\tilde{G}_{\rm{diag}}\,,
\eeq
where $N'$ and $N_{\rm{diag}}$ are the group factors, 
\beq
N'=N_{\rm{diag}}=\sum_{LH-RH}Tr\left[T_{PQ}^a\{t^b,t^c\}\right]=12\,, 
\eeq
and where $T_{PQ}^a$ corresponds to the PQ generator and $t^b=\frac{\lambda^b}{2}$ to the Gell-Mann matrices for the $SU(3)$ generators. 
 The anomalous term modifies the classical equations of motion of the axion,
\beq
f_a \,\Box a +\partial_\mu(\overline{q'_L}\gamup q'_L)+\partial_\mu(\overline{u_L'^c}\gamup u_L'^c)+\partial_\mu(\overline{d_L'^c}\gamup d_L'^c)=\frac{\alpha_{\rm{diag}}}{8\pi}12\,G_{\rm{diag}}\tilde{G}_{\rm{diag}}\,, 
\eeq
and gives rise to an effective Lagrangian, 
\beq
\mathcal{L}\supset \frac{1}{2}\partial_\mu a\partial^\mu a+\frac{1}{f_{a}}\partial_\mu a\,\left(\overline{q'_L}\gamup q'_L+\overline{u_L'^c}\gamup u_L'^c+\overline{d_L'^c}\gamup d_L'^c \right)+\frac{\alpha_{\rm{diag}}}{8\pi}\,\frac{12}{f_a}\,a\,G_{\rm{diag}}\tilde{G}_{\rm{diag}}\,.
\label{effective-axion}
\eeq
The impact of the SSI of the spontaneously broken theory will again add further contributions, inducing a putatively high mass for the elementary axion as discussed further below. 

\begin{table}[t]
\begin{align*}
\begin{array}{c|c|c}
	& SU(3)_c	& SU(3)_{\rm{diag}}	\\
\hline
\psi_{L}	& \Box		& \bar\Box	\\
\psi_{L}\,^c	& \bar\Box		& \Box	
\end{array}
\end{align*}
\caption{The massless quark sector charged under $SU(3)_{\rm{diag}}$ below $\Lambda_{\rm{CUT}}$ in the model with an extra scalar.}
\label{massless-below P}
\end{table}

\subsubsection{Running of the coupling constants}
 The matter content allows $SU(3)_{\rm{diag}}$ to confine at higher scales than the QCD group $SU(3)_c$ as in model I. The separation of both scales is made even sharper in the model II because $\alpha_{\rm{diag}}$ runs faster. In model II, only one massless fermion charged under $SU(3)_{\rm{diag}}$ is present under the CUT scale (compare Tabs.~\ref{massless-below-chi} and \ref{massless-below P}).
\begin{figure}[t]
\centering
\includegraphics[scale=.65]{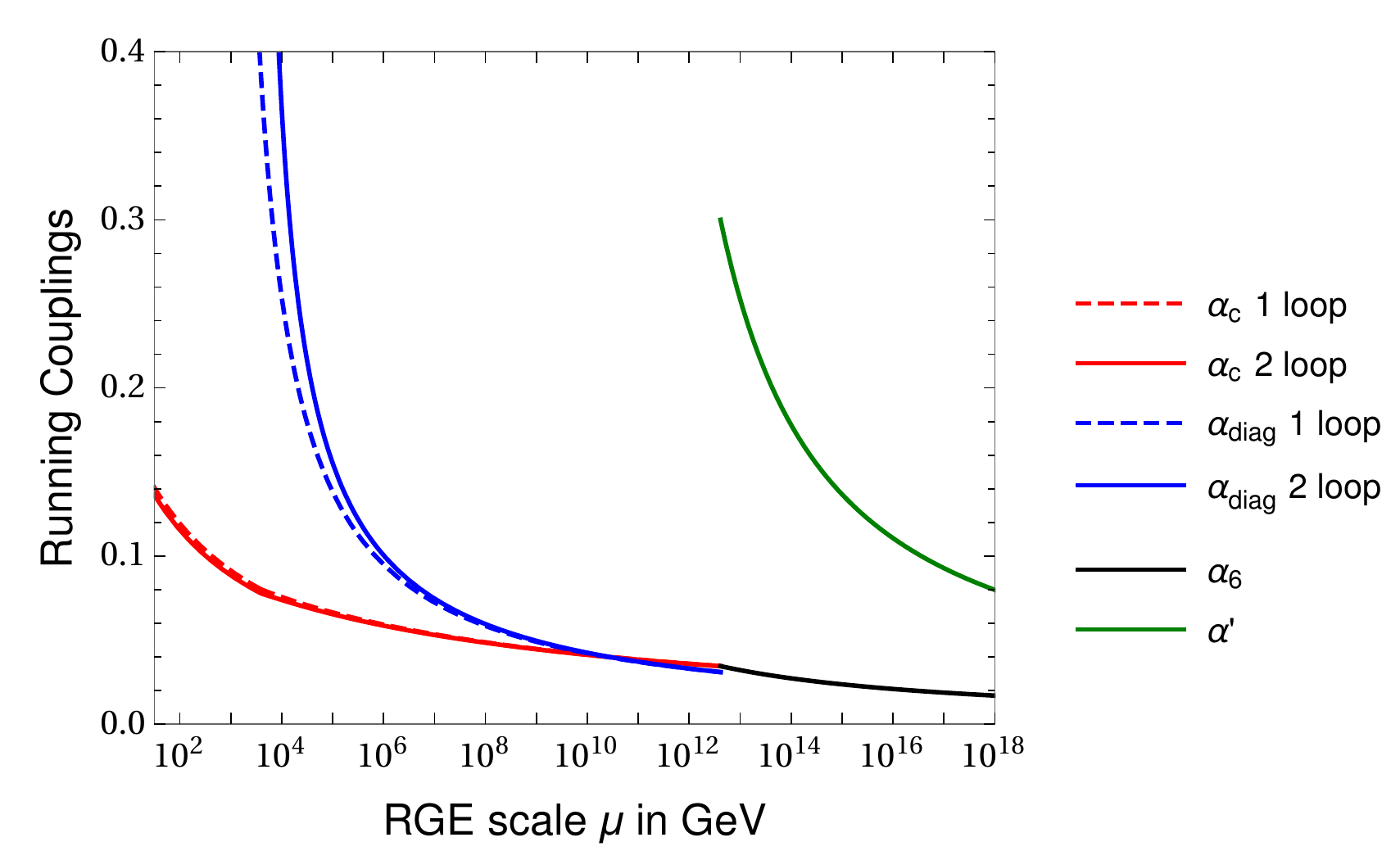}
\caption{Running of $\alpha_{\text{QCD}}$, $\alpha_{\text{diagonal}}$, 
$\alpha_6$, and $\alpha'$ in the model with one extra $SU(6)$ scalar; the only inputs assumed are  
$\alpha'_{\rm{CUT}} = 0.3$ and $\Lambda_{\rm{CUT}}= 4.1\times10^{12}$  
GeV for illustration, which results in $\Lambda_{\rm{diag}}= 4$ TeV (taken as a benchmark point). The solid (dashed) lines correspond to the two (one) loop results.
 } 
\label{Running-Psi}
\end{figure}
We have estimated both the one and two-loop running, as illustrated in Fig.~\ref{Running-Psi} for $\Lambda_{\rm{diag}}=4$ TeV. 

\subsubsection{Confinement of $SU(3)_{\rm{diag}}$ and pseudoscalar anomalous couplings to the confining interactions.}
At the scale $\Lambda_{\rm{diag}}$ the QCD coupling constant is small, and the $SU(3)_{\rm{diag}}$ spectrum with only one massless fermion in Tab.~\ref{massless-below P}  has an approximate  classical global symmetry $U(3)_L \times U(3)_R$. Upon chiral symmetry breaking $U(3)_L \times U(3)_R\longrightarrow U(3)_V$  by the  quark condensate $\langle\bar\psi_L \psi_R\rangle$, nine pGBs appear,
\begin{equation}
9 = 1_c \,+\, 8_c\,.
\label{9-decomp}
\end{equation}
 The gauge QCD group $SU(3)_c$ is again a subgroup of $U(3)_V$ which remains unbroken. 
 The octet of pGBs  colored under QCD  will acquire large masses due to gluon loops, which at one-loop is given by 
\begin{align}
m^2(8_c) 
	&\approx \frac{9\,\alpha_\text{c}}{4\pi} \Lambda_\text{diag}^2\,.
	\label{colored-1}
\end{align} 
The QCD singlet $1_c$, denoted $\eta'_\psi$,  is a dynamical axion. Note that it has the same quark composition as the $\eta'_\psi$ meson in model I. The $\eta'_\psi$ couples to both the $SU(3)_{\rm{diag}}$ and $SU(3)_c$ anomalies,
\beq
j^\mu_{\psi_A}=\bar{\psi}\gamup\gamma^5 t^9 \psi \equiv f_{\rm{d}}\,\partial^\mu\eta'_\psi\, , \qquad 
t^9=\frac{1}{\sqrt{6}}\mathds{1}_{3\times3}\\,,
\label{jetaprimed}
\eeq
resulting in a low-energy effective Lagrangian for this axion given by
\beq
\mathcal{L}_{eff}\subset - \frac{\sqrt{6}\,\eta'_\psi}{f_{\rm{d}}}\,\left (\frac{\alpha_{s}}{8\pi}\,G_{c}\tilde{G}_{c}+\frac{\alpha_{\rm{diag}}}{8\pi}\,G_{\rm{diag}}\tilde{G}_{\rm{diag}}\right)\,.
\eeq

	In summary, this solution to the strong CP problem is a hybrid one  with two axions: a heavy dynamical axion $\eta'_\psi$ with mass of order $\Lambda_{\rm{diag}}$  stemming from a PQ symmetry which reabsorbs the original $\theta_{SU(6)}$ (and thus  $\theta_{\rm{QCD}}$) parameter as in the previous section,  and a second elementary axion $a$ resulting from solving the external $SU(3')$ sector {\it \`a la} PQWW~\cite{Peccei:1977hh,Weinberg:1977ma,Wilczek:1977pj}.  Up to now, only two sources of masses have been identified for the ensemble of three pseudoscalars with anomalous couplings ($\eta_{\rm{QCD}}'$, $\eta'_\psi$ and $a$). 
	We analyze next the SSI of this model which provide a large source of axion mass for the elementary axion $a$.

\subsubsection{Impact of small-size instantons of the spontaneously broken CUT}
\label{sec-SSI-psi}

\begin{figure}[t]
\centering
\begin{centering}
 \includegraphics[scale=.85]{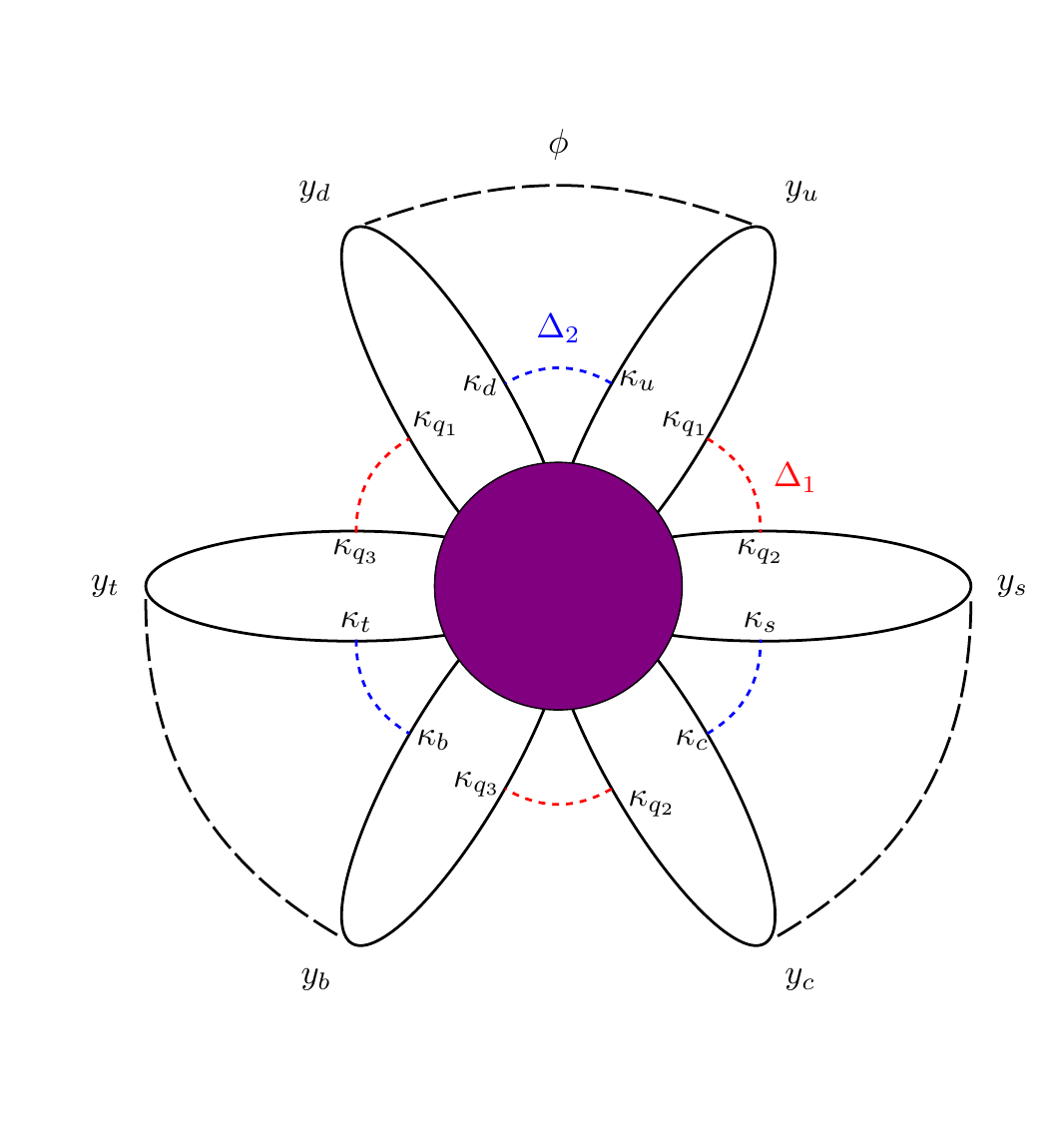} 
\caption{Instanton contribution in model II. 
The long dashed lines connecting the SM Yukawa interactions correspond to $\phi$ propagators while the short dashed lines depict $\Delta_1$ or $\Delta_2$ propagators.  
} \label{instantons_model2}
\end{centering}
\end{figure}

 The analysis of SSI for model II under discussion is simpler than that for model I 
 developed in the previous subsection.  {\it No} massless fermions charged under $SU(3')$  are present in model II (in contrast with model I). Furthermore, 
PQ symmetry forbids here the $y_i'$ Yukawa couplings which gave the dominant contribution in model I. In consequence, the terms proportional to $\kappa_i$ and mediated by the $\Delta_1$ and $\Delta_2$ scalars  and the Higgs field will dominate $\Lambda_{\rm{SSI}}$. This is illustrated by the instanton ``flower'' in Fig.~\ref{instantons_model2}. It results in  
   $\Lambda_{\rm{SSI}}$  given by 
    	\beq
\Lambda_{\rm{SSI}}^4 =  -\int \frac{d\rho}{\rho^5}\, D[\alpha'(1/\rho)] \frac{1}{(4\pi)^{18}} \prod_i  Y^{SM}_{\,u_i} Y^{SM}_{\,d_i}\,(\kappa_q^i)^2 \kappa_u^i \kappa_d^i \,,
 \eeq
 which can be written as  
 \beq
\Lambda_{\rm{SSI}}^4=  C_{inst}\Lambda_{\rm{CUT}}^{b}e^{-\frac{2\pi}{\alpha'_{\rm{CUT}}}}\frac{1}{(4\pi)^{18}}\prod_i  Y^{SM}_{\,u_i}Y^{SM}_{\,d_i} \,\kappa_q^{i\,^2} \kappa_u^i \kappa_d^i  \int_0^{1/\Lambda_{\rm{CUT}}} d\rho\, \rho^{b-5} \left(-b\,\ln\left(\rho\Lambda_{\rm{CUT}}\right)+\frac{2\pi}{\alpha'(\Lambda_{\rm{CUT}})} \right)^6 \,.
 \eeq
 with $b=5$ in this case. Here the approximation 
in Eq.~\eqref{Approx} 
is no longer valid, and the result corresponds to that of the pure Yang-Mills case (Eq.~\eqref{YangResult}) with the extra suppression factor of the Yukawa couplings,
 \beq
\Lambda_{\rm{SSI}}^4 = C_{inst}\,f(\alpha'_{\rm{CUT}},b)\,e^{-\frac{2 \pi }{\alpha'_{\rm{CUT}}}}  \Lambda_{\rm{CUT}}^4\,\frac{1}{(4\pi)^{18}} \prod_i Y^{SM}_{\,u_i} Y^{SM}_{\,d_i}\,\kappa_q^{i\,^2} \kappa_u^i \kappa_d^i 
 \eeq
 where the function $f(\alpha'_{\rm{CUT}},b)$ is defined in Eq.~\eqref{ffunction}.
 
\begin{figure}[t]
\centering
\includegraphics[scale=.55]{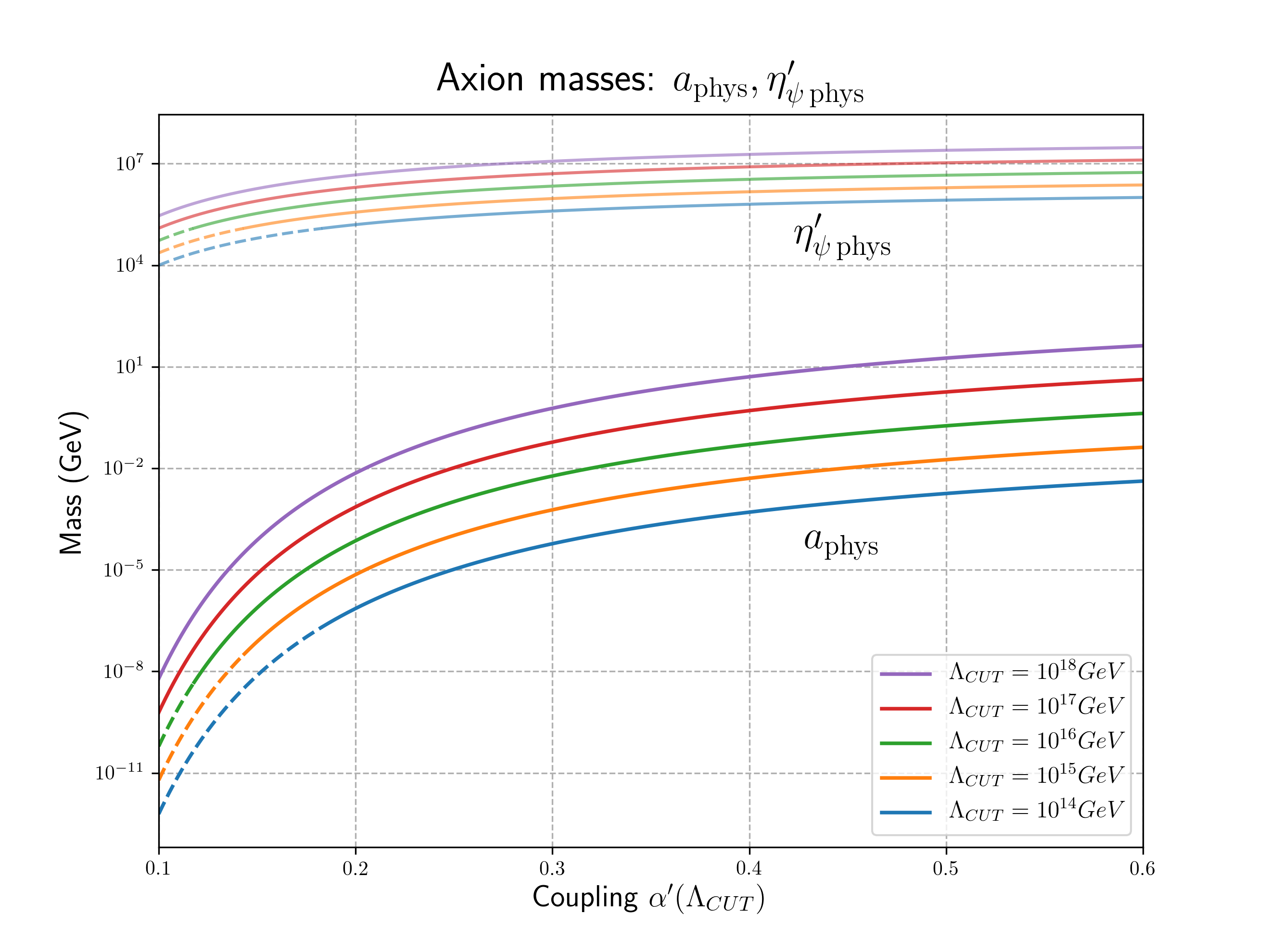}
\caption{Model with a second $SU(6)$ scalar.  The $a$ axion mass induced by the $SU(3')$ small size instantons  at the scale $\Lambda_{\rm{CUT}}$ is illustrated, together with the larger mass for the dynamical axion $\eta'_\psi$ sourced by $SU(3)_{\rm{diag}}$ instantons. The $\alpha_{\rm{CUT}}'$ values allowed correspond to the solid sectors of the lines, while dashed lines correspond to excluded values, for which $SU(3)_{\rm{diag}}$ would confine below  $2.9$ TeV, ruled out by searches for scalar octets at LHC~\cite{Aaboud:2017nmi} (see Sec.~\ref{collider}).  }
\label{LambdaSSI-psi}
\end{figure}

 This result translates into a new contribution to the  instanton-induced effective potential of the form  
\begin{equation}
 \delta \mathcal{L}_{eff}=\Lambda_{\rm{SSI}}^ 4\,\cos\left(12\,\frac{a}{f_a}\right)\,.
 \label{SSI-psi}
\end{equation}
Taking into account that in this model the elementary axion scale coincides with the CUT scale, it follows that for $\alpha'_{\rm{CUT}}=0.3 $,
\begin{equation}
m_a^2 \sim  3.7 \times 10^{-37}\Lambda_{\rm{CUT}}^2\,.
\end{equation}
The linear dependence  of $m_a$ on $\Lambda_{\rm{CUT}}$  implies that this second axion is generically much heavier than the corresponding one in model I (e.g. Eq.~(\ref{meta'-SSI-chi})), see 
 Fig.~(\ref{LambdaSSI-psi}) for illustration.  In summary, the dynamical axion has  a mass of order $\Lambda_{\rm{diag}}$ and thus  of a few TeV or above, while the elementary axion is generically extremely heavy (with again the caveat in case of Yukawa fine tunings discussed in Sec.~\ref{sec-SSI}).  As in model I, no light axion remains.

\subsubsection{Solution to the strong CP problem.}
 The minimum of the axion potential can be easily shown to remain CP-conserving after including all contributions to the axion masses. Indeed, the $\theta_i$ dependence of the Lagrangian can be again read off of Eq.~\eqref{twotheta},
\begin{align}
\mathcal{L}=\Lambda_{\rm{SSI}}'^ 4\,\cos\left(12\,\frac{a}{f_{a}}+\bar\theta'\right)\,+\Lambda_{\rm{d}}^ 4\,\cos\left(12\,\frac{a}{f_{a}}-\sqrt{6}\,\frac{\eta'_{\psi}}{f_{\rm{d}}}+\bar\theta'+\bar\theta_6\right)+\Lambda_{\rm{QCD}}^ 4\,\cos\left(-\sqrt{6}\,\frac{\eta'_{\psi}}{f_{\rm{d}}}+\bar\theta_6\right)\,,
\end{align}
 for which the following bosonic vevs lead to a CP-conserving minimum,
 \begin{align}
\left\langle12\,\frac{a}{f_{a}}+\bar\theta'\right\rangle=0 \,,&& 
\left\langle\bar{\theta}_{6}-\,6\,\frac{\eta'_{\psi}}{f_{\rm{d}}}
\right\rangle=0\,.
\label{thetaminima}
\end{align} 
We recall once again that computing the exact dependence of the potential on the phases of the Yukawa couplings is a task which remains for future work. After the replacement $a\to \langle a\rangle + a$, $\eta'_{\psi} \to \langle\eta'_{\psi}\rangle +\eta'_{\psi} $ and  introducing as well the QCD $\eta'_{\rm{QCD}}$ field, the effective low-energy mass Lagrangian for the physical mesons which couple to anomalous currents is given by
 \begin{align}
\mathcal{L}=\Lambda_{\rm{SSI}}^ 4\,\cos\left(12\,\frac{a}{f_{a}}\right)\,+\Lambda_{\rm{d}}^ 4\,\cos\left(12\,\frac{a}{f_{a}}-\sqrt{6}\,\frac{\eta'_{\psi}}{f_{\rm{d}}}\right)+\Lambda_{\rm{QCD}}^ 4\,\cos\left(2\,\frac{\eta'_{\rm{QCD}}}{f_\pi}+\sqrt{6}\,\frac{\eta'_{\psi}}{f_{\rm{d}}}\right)\,.
\label{finalpotential2}
\end{align}
where all the CP violating phases have been relaxed to zero.

\subsubsection{Computation of the pseudoscalar mass matrix: $a$,  $\eta'_\psi$, $\eta'_{\rm{QCD}}$ and light spectrum}
Taking into account all contributions except the SM quark masses, the following mass matrix results for the singlet pseudoscalars of the theory which couple to anomalous currents:  
\beq
M^2_{a,\,\eta'_\psi,\,\eta'_{\rm{QCD}}}=\,\left(\begin{array}{ccc}
144\,\frac{\left(\Lambda_{\rm{SSI}}^ 4+\Lambda_{\rm{diag}}^ 4\right)}{f_{a}^2}&-12\sqrt{6}\,\frac{\Lambda_{d}^ 4}{f_{\rm{d}}f_{a}}&0\\ 
-12\sqrt{6}\,\frac{\Lambda_{d}^ 4}{f_{\rm{d}}f_{a}} & 6\,\frac{\left(\Lambda_{d}^ 4+\Lambda_{\rm{QCD}}^ 4\right)}{f_{\rm{d}}^2} & 2\sqrt{6}\,\frac{\Lambda_{\rm{QCD}}^ 4}{f_{\pi}f_{\rm{d}}}\\ 
0&2\sqrt{6}\,\frac{\Lambda_{\rm{QCD}}^ 4}{f_{\pi}f_{\rm{d}}}& 4\frac{\Lambda_{\rm{QCD}}^ 4}{f_{\pi}^2}
\end{array}\right)\,.
\label{massmat-1}
\eeq
Assuming the hierarchy of scales $\frac{\Lambda_{d}^ 4}{f_{\rm{d}}^2}\gg\frac{\Lambda_{\rm{QCD}}^ 4}{f_\pi^2}\gg\frac{\Lambda_{\rm{SSI}}^ 4}{f_{a}^2}$, the resulting mass eigenvalues  are 
\begin{align}
m^2_{a_{phys}}\simeq144\frac{\Lambda_{\rm{SSI}}^ 4}{f_{a}^2}\,,\qquad
m^2_{\eta'_{\psi,\,phys}}\simeq6\frac{\Lambda_{d}^ 4}{f_{\rm{d}}^2}\,,\qquad
m^2_{\eta'_{QCD\,phys}}\simeq4\frac{\Lambda_{\rm{QCD}}^ 4}{f_\pi^2}\,.
\label{pseudoscalarmasses-II}
\end{align}
In this model both, the usual QCD $\eta'$ and the axion $a_{phys}$ remain as light eigenstates, while the other eigenstate $\eta'_{\psi,\,phys}$ will have a mass generically above the TeV scale. 

Apart from the axion, the lowest set of exotic states is an octet of exotic ``pions'' whose masses are $\lesssim$ TeV, see Eq.~(\ref{colored-1}). A similar octet is present in model I discussed in Sec.~\ref{3mesons}, but model I contains an additional color-triplet set of pseudoscalars. No color-triplet is expected here as the exotic classical flavor symmetry is $U(3)_L\times U(3)_R$, see Eq.~\eqref{9-decomp}, instead of the $U(4)_L\times U(4)_R$ symmetry of model I.

\vspace{1cm}

This model II with an additional scalar may be less appealing than than model I with an extra massless fermions  for two reasons: a)  its axion sector contributes directly to the EW hierarchy problem, as its elementary axion results from a scalar potential which a priori communicates with the Higgs potential; b) it is a hybrid model with both one elementary and one dynamical axion, while model I is more aligned with the spirit of solving fully the strong CP problem via massless fermions. 

\section{Phenomenological and cosmological  limits on the lightest exotic states}
\label{pheno}

A common feature of both ultraviolet complete models constructed above is that
 the generic spectrum under the EW scale is the SM spectrum plus sterile fermions,
  in contrast with usual axion models.
\subsection{Collider observable signals}
\label{collider}
The lowest set of observable exotic states 
 are expected to be the exotic $SU(3)_c$-colored ``pions''  whose masses may lie under the TeV scale. 
These resulted from the chiral symmetry breaking of the confining group $SU(3)_{\rm{diag}}$.  In model I, QCD color-triplet and color-octet bound states are made out of $\psi$ and $\chi$ massless fermions, shown in Tab~\ref{massless-below-chi} and Eq.~(\ref{colored-2}). Model II gives only color octets composed of $\psi$ fermions, as shown in Tab.~\ref{massless-below P} and Eq.~(\ref{colored-1}).

 In this color-unified axion solution, the exotic fundamental fermions have no SM $SU(2)\times U(1)$ charges. The heavy pions will be  produced in colliders  only via QCD interactions, e.g. gluon-gluon couplings, through which they also presumably decay before they can hadronize to make colour neutral states. As they are coloured, they do not mix with ordinary pions or other visible matter. 

\subsubsection{Color-octet pions from $SU(3)_{\rm{diag}}$ confinement}
The lightest scalar octets in Eqs.~(\ref{colored-2}) and (\ref{colored-1}) are denoted by $\pi_{\rm{d}}\equiv 8_c$. Their effective coupling to QCD gluons can be written as 
\[ \label{pid_lag}
\mathcal{L}\supset D_\mu \pi_{\rm{d}} D^\mu \pi_{\rm{d}} + \frac{3\sqrt{3}\,\alpha_s}{8\pi}\frac{\pi_{\rm{d}}^a}{f_d}  d_{abc}G_{\mu\nu}^b\tilde{G}^{c\mu\nu}\,,
\]
where $D_\mu$ denotes the  $SU(3)_c$ covariant derivative  and 
  $d_{abc}$ is the corresponding symmetric group structure constant. The second term in Eq.~(\ref{pid_lag}) results at one-loop from the triangle diagram with the fermions $\chi$ running in the loop.  The color-octet exotic pions can thus be either produced in pairs through the gluon-gluon-$\pi_{\rm{d}}$-$\pi_{\rm{d}}$ coupling in the kinetic term, or 
  singly produced through their anomalous coupling to gluons. The kinetic term dominates the scalar octet production channels, while the second term allows the $\pi_{\rm{d}}$ decay, yielding a dijet final state.

Experimental limits on scalar octet pair production via the gluonic interactions in the kinetic term can be inferred by recasting searches of sgluons.
A recent search of sgluon pair production by ATLAS using $\unit[36.7]{fb^{-1}}$ of $\sqrt{s}=\unit[13]{TeV}$ data~\cite{Aaboud:2017nmi}, whose prediction was obtained from the NLO computation in Ref.~\cite{GoncalvesNetto:2012nt} for $\sqrt{s}=8$ TeV (rescaled to $\unit[13]{TeV}$ according to Ref.~\cite{Degrande:2014sta}), sets a bound on the octet scalar exotic pions given by
\begin{equation}
m(\pi_{\rm{d}}) \gtrsim\unit[770]{GeV}\,.
\end{equation}
From Eqs.~(\ref{colored-2}) and (\ref{colored-1}), this translates into 
\begin{equation}
\Lambda_{\rm{diag}} \gtrsim\unit[2.9]{TeV}\,.
\end{equation}
For model I, an alternative bound may be inferred from the limits on color-triplet scalars, which can be produced via their color interactions. In the absence of couplings which mediate their decay (as the second type of coupling in Eq.~(\ref{pid_lag}) is not possible for scalar triplets), they will bind with SM quarks to form stable hadrons. This search is expected to result in a sensitivity similar to the one above~\cite{Aaboud:2016uth}.

It is very interesting to pursue the experimental search for colored pseudoscalars and stable exotic hadrons. Their detection would be a powerful indication of the dynamical solution to the strong CP problem proposed here.

\subsubsection{ Dynamical axion and exotic fermions}
The dynamical axion denoted above by  $\eta'_{\psi,\,phys}$ with instanton-induced mass  of order $\Lambda_{\rm{diag}}$, Eqs.~(\ref{pseudoscalarmasses-I}) and (\ref{pseudoscalarmasses-II}), 
can {\it a priori}  be either pair-produced through the kinetic coupling or singly produced through the dimension five anomalous operator. It would decay dominantly to two back-to-back jets and can be searched for in dijet resonance searches.   Its production, however, may be suppressed by its high mass.  For instance, for $\Lambda_{\rm{diag}} \simeq \unit[2.9]{TeV}$, $m_{\eta'_{\psi,\,phys}} \sim \unit[90]{TeV}$, which is beyond the reach of present collider searches. 
 The other mesons and baryons resulting from the $SU(3)_{\rm{diag}}$ confinement have masses in the TeV range and above; they would lead to collider signatures similar to those for the exotic pseudo-goldstone bosons. 

\subsubsection{The axion coupled to SSI}
In model I with its additional massless fermion, this second axion is also dynamical and was denoted by $\eta'_{\chi,\,phys}$. For the benchmark case with Yukawa couplings in the  prime-fermion sector of order one, its mass is expected to lie above that of the $\eta'_{\psi,\,phys}$ and out of reach, as shown in Eq.~(\ref{pseudoscalarmasses-I}). As discussed in Sec.~\ref{sec-SSI}, it is possible for the prime-sector fermions to have smaller Yukawa couplings. The SSI-induced scale which sources this axion mass would then be substantially lowered, resulting in an $\eta'_{\chi,\,phys}$ mass similar to or much smaller than the $\eta'_{\psi,\,phys}$ mass. In fact, the $\eta'_{\chi,\,phys}$  could even become the lightest exotic state observable at colliders, given that its couplings to the visible world are only suppressed by $f_{\rm{d}}$.  A dedicated study would be required to examine how light this axion can become while maintaining this realistic ultraviolet model.

The second axion in model II, denoted $a$ and of elementary nature, could analogously become very light by fine-tuning the prime-fermion Yukawa structure. Nevertheless, as its associated axion scale is the color-unified scale, $f_a\sim \Lambda_{\rm{CUT}}$, its couplings to the visible world are extremely suppressed. It follows that such light axion would not be accessible at present or foreseen colliders.

 \subsection{Cosmological and Gravitational aspects} 
 
We briefly discuss next the cosmological aspects of the models constructed above, as well as the putative instability threat from gravitational non-perturbative effects. 
 
 \subsubsection{ Stable particles and cosmological structures}
 
 Stable particles with masses higher than about $10^5$ GeV may lead to cosmological problems, dominating the mass density and overclosing the universe~\cite{Kolb:1990vq}. This is often a problem in previous models of composite axions because exotic stable baryons  bound by the extra confining force~\cite{Choi:1985cb, Choi:1985qu} are expected.
 
 However, as pointed out in Refs.~\cite{Choi:1985qu, Dobrescu:1996jp}, if axicolor can be unified with a SM gauge
group, then the unified forces could mediate the decay of axihadrons into
lighter states, and the model would be cosmologically safe.  The color unification of this work automatically employs this mechanism. The heavy exotic hadrons decay
 to the sterile fermions $\psi_\nu$, which are part of the CUT massless multiplet $\Psi$ in Eq.~(\ref{Psidecomposition}) and Tab.~\ref{massless-fermions}. It remains to be determined whether the lifetime for these CUT-induced decay channels is too large to avoid problems from stable exotic hadrons. If the decays are made to be fast enough, the resulting sterile fermions may induce in turn other cosmological problems. This analysis is left to a future work.
 
 A similar concern pertains to the domain walls  which may arise due to the spontaneous breaking of the discrete shift symmetry in the instanton potential, Eqs.~(\ref{axionpotential-2}) and (\ref{finalpotential2}). The walls need to disappear before
they dominate the matter density of the universe, or else other mechanisms must be applied to  solve the domain wall problem~\cite{Cheng:1987gp, Barr:1992qq, Preskill:1991kd, Dvali:1995cc, Kim:1984pt}. 

In any case, 
if the universe went through an inflation phase  any relic previously present will be wiped out.  If the reheating temperature is
lower than the PQ scales, meaning lower than $\Lambda_{\rm{diag}}$ here, neither the heavy stable particles nor any putative domain walls are produced again after inflation, and the problems mentioned above are avoided altogether. 
   We defer to a future work the in-depth study of the cosmological aspects of model I and model II, with either high- or low-scale inflation.

\subsubsection{The question of gravitational quantum effects}

Gravitational quantum corrections have been suggested to be relevant
and dangerous for axion models in which the PQ scale is not far from the Planck scale.  In model I, both PQ scales correspond to $\Lambda_{\rm{diag}}$, which is much lower than the Planck scale, and no instability resulting
from gravitational quantum effects is at stake.  In model II instead,
while the dynamical PQ scale is analogously low, the second PQ
symmetry is realized at the CUT scale and gravitational quantum
effects could be relevant.

It has often been argued that all global symmetries may be violated by non-perturbative quantum gravitational effects, see for instance Refs.~\cite{Holman:1992us,Kamionkowski:1992mf, Barr:1992qq, Ghigna:1992iv, Georgi:1981pu}. For instance, a black hole can eat global charges and subsequently evaporate. Similar effects may exist with virtual black holes.   Another indication that gravity might not respect global symmetries comes from
wormhole physics~\cite{Giddings:1988cx,Coleman:1988tj,Gilbert:1989nq, Rey:1989mg}. The natural scale of violation in this case is the wormhole scale, usually thought to be very near (within an order of magnitude or so) the Planck mass $M_{\rm{Planck}}$.  

For axion models with high PQ scales, such as the typical scale of invisible axion models  $f_a\sim 10^{9}-10^{12}$ GeV, it has been argued that those non-perturbative quantum gravitational effects could lead to extreme fine-tunings.  The authors of Ref.~\cite{Holman:1992us,Kamionkowski:1992mf, Barr:1992qq, Ghigna:1992iv} concentrated on the simplest (and the most dangerous) hypothetical dimension five effective operator 
\begin{equation}
g_5\, \frac {|\Phi|^4\, (\Phi+ \Phi^*) }{M_{\rm{Planck}}} \,,
\label{gravity-5}
\end{equation}
where $g_5$ is a dimensionless coefficient and $\Phi$ would be a field whose vev breaks the PQ invariance. 
 This term threatens the standard invisible axion solutions to the strong CP problem, as it  would change the shape of the effective potential. The minimum moves unacceptably away from  a CP-conserving solution unless the coefficient is strongly fine-tuned, for instance $g_5< 10^{-54}$ for $f_a\sim 10^{12}$ GeV.  
  These potentially dangerous terms can be avoided if the global PQ symmetry arises accidentaly as a consequence of other gauge groups \cite{Randall:1992ut, Redi:2016esr}. Nevertheless, the idea that gravity breaks all global symmetries is indeed an assumption\,---\,and sometimes an
incorrect one\,---\,at least at the Lagrangian level.~\footnote{For example, orbifold compactifications of the
heterotic string have discrete
symmetries that prevent the presence of some higher dimension operators, and  this can strongly  and safely 
suppress the dangerous  effects under discussion~\cite{Butter:2005wr}.}
  Furthermore, very recently the impact of  non-perturbative effects  has been clarified and quantified in Ref.~\cite{Alonso:2017avz}.  The effects turn out to be extremely suppressed by an exponential dependence on the gravitational instanton action, and they are harmless even with high axion scales. The demonstration relied only on assuming that the spontaneous breaking of the PQ symmetry is implemented through the vev of a scalar field, and thus it directly applies to our model II.\footnote{ Although no explicit demonstration was given in that work for  the case of dynamical breaking via condensates,  plausibly the result would also apply for models with dynamical axions and very high axion scales.} In summary, both model I and model II are safe from instabilities induced by gravitational quantum corrections.

\section{Conclusions}

Color unification with massless quarks has been proposed and developed here for the first time.  As a simple implementation of the idea, the SM color group has been embedded in $SU(6)$, which is spontaneously broken to QCD and a second confining and unbroken gauge group. An exactly massless $SU(6)$ fermion multiplet solves the strong CP problem. We have  fully developed two ultraviolet completions of the mechanism.

In order to implement this idea successfully, it is necessary to give satisfactorily high masses to the $SU(6)$ partners of the SM quarks to achieve a separation between the QCD scale and that of the second confining group.  For this purpose, 
an auxiliary $SU(3')$ gauge group is introduced under which the aforementioned massless fermion is a singlet. $SU(6)\times SU(3')\to SU(3)_c\times SU(3)_{\rm{diag}}$ is a simple and realistic option. 
Both final groups 
 remain unbroken and confine at two different scales, $\Lambda_{\rm{QCD}}$ and $\Lambda_{\rm{diag}}$, with $\Lambda_{\rm{diag}}\sim \mathcal{O}(\# $ TeV$) \gg \Lambda_{\rm{QCD}}$.  
The scale $\Lambda_{\rm{diag}}$ then gives the order of magnitude of the mass of the dynamical composite axion inherent to the color-unified mechanism. 
Furthermore, massless (or almost massless) sterile fermions are  a low-energy trademark remnant of the massless multiplet that solves the SM strong CP problem.

 In order to avoid 
the   $SU(3')$ sector sourcing back an extra contribution to the strong CP problem, a minimal extension of its matter sector suffices. Two examples of ultraviolet complete models have been explored in this work: in model I an extra $SU(3')$ massless fermion is added, while  model II includes instead a second scalar with the same quantum numbers as the color-unification breaking scalar. From the point of view of the strong CP problem, those two models are very different. Model I features a second dynamical axion with a second PQ scale of order $\Lambda_{\rm{diag}}$. In model II, this second PQ scale coincides with the much larger color-unification scale, and the associated axion is elementary.  We computed the two-loop running of all coupling constants involved, showing that the desired separation of all relevant scales is achieved naturally: a color-unification scale much larger than the two confining ones, $\Lambda_{\rm{diag}}$ and $\Lambda_{\rm{QCD}}$, and the subsequent separation of the last two.  This separation of scales is robust and stable over a wide range of parameter values.

We have found that regardless of the details of the ultraviolet implementation, generically there are the three sources of anomalous currents:  the instantons of the confining $SU(3)_c$, the instantons of the confining $SU(3)_{\rm{diag}}$, and finally the small-size instantons of the spontaneously broken color-unified theory. There are thus three diverse sources of mass for the three pseudoscalars in the theory which couple to anomalous currents: 
the QCD $\eta'$, the dynamical axion inherent to color unification, and the second axion (either dynamical or elementary) associated to the solution of the $\theta'$ problem. These three bosons then acquire masses of order $\Lambda_{\rm{QCD}}$, $\Lambda_{\rm{diag}}$ and $\Lambda_{\rm{SSI}}$, respectively, 
 and no standard invisible axion is left in the low-energy spectrum. This is generically a very interesting mechanism from the point of view of solving the strong CP problem with heavy axions and scales around the TeV. The mechanism allows a wide extension beyond the invisible axion range of axion parameter space which solves the strong CP-problem. 
 
 With axion masses and scales around the TeV, observable signals at colliders are expected, as well as other rich phenomenology.   
  Generically, the lightest exotic bound states are colored pseudoscalars (QCD octets and triplets in model I, and only octets in model II).  We have recast the results from present experimental searches of heavy colored mesons to infer a $\unit[2.9]{TeV}$ bound on the confinement scale of the second confining group, $SU(3)_{\rm{diag}}$, which is directly related in model I to the axion scale. 
  It is possible, however, that the lightest hadron is one of the axions instead of a colored meson, although this possibility has been shown to require some fine-tuning of the Yukawa couplings of the theory and is consequently less appealing.   

 Overall, model I may be preferred as: i) it is exclusively based on solving the strong CP problem dynamically  via massless quarks; ii) from the point of view of naturalness it does not require any fine-tuning to ensure the hierarchy between the PQ and the electroweak scales, as no PQ field is involved in the scalar potential. In model II instead, one PQ field participates in the color-unification scalar potential,  and furthermore this model is a hybrid dynamical-elementary axion solution to the strong CP problem.
 
Model I is also unquestionably safe from the point of view of stability of the axion solution with respect to non-perturbative effects of quantum gravity, as its PQ scales are $\Lambda_{\rm{diag}}\sim$ TeV. Furthermore, recent advances suggest that the quantum gravity threat should not be considered a risk even for model II. The other issue of the cosmological impact of (quasi) stable heavy exotic hadrons and of the (almost) massless sterile fermion remnants can  be simply avoided by introducing an inflation scale and reheating temperature lower than $\Lambda_{\rm{diag}}$. This last subject deserves future detailed attention in particular in view of the dark matter puzzle.

\section*{Acknowledgments}

We acknowledge M. Garc\'ia P\'erez, T. Yanagida, V. Sanz, K. Harigaya, and Simon Knapen for very interesting conversations and comments. 
M.B.G, R. H., R.dR  and P. Q. acknowledge Berkeley LBNL, where part of this work has been developed.
This project has received funding from the European Union's Horizon 2020 research and innovation programme under the Marie Sklodowska-Curie grant agreements No 690575  (RISE InvisiblesPlus) and  No 674896 (ITN ELUSIVES). M.B.G, R. H., R.dR  and P. Q. also acknowledge support from the 
 the Spanish Research Agency (Agencia Estatal de Investigaci\'on) through the grant IFT Centro de Excelencia Severo Ochoa SEV-2016-0597. M.B.G,  R.dR  and P. Q. acknowledge as well support 
 from the "Spanish Agencia Estatal de Investigaci\'on" (AEI) and the EU ``Fondo Europeo de Desarrollo Regional'' (FEDER) through the project FPA2016-78645-P. 
  The work of M.K.G. was supported in part by the Director, Office of
Science, Office of High Energy and Nuclear Physics, Division of High
Energy Physics, of the U.S. Department of Energy under Contract
DE-AC02-05CH11231, and in part by the National Science Foundation under
grant PHY-1316783. The work of  R. H. was supported to and ESR contract of the H2020 ITN Elusives. The work of P.Q. was supported through a ``La Caixa-Severo Ochoa'' predoctoral grant of Fundaci\'on La Caixa.


\providecommand{\href}[2]{#2}\begingroup\raggedright\endgroup

\end{document}